\documentclass[]{tandf2_AOM}
\usepackage{graphicx}
\usepackage{amssymb}
\usepackage{amsthm}
\usepackage{amsmath}
\usepackage{mathtools}
\newcommand{\p}[2]{\frac{\partial #1}{\partial #2}}

\newcommand{\pl}[2]{\partial #1/\partial #2}

\newcommand{\dl}[2]{{\rm d} #1/{\rm d} #2}

\newcommand{\dd}[2]{\frac{{\rm d}^2 #1}{{\rm d} #2^2}}
\newcommand{\der}[2]{\frac{{\rm d} #1}{{\rm d} #2}}
\usepackage{float}    
\usepackage{ctable} 
\usepackage{booktabs} 
\usepackage{psfrag}  
\usepackage{xcolor}
\usepackage{caption}

\theoremstyle{plain}

\theoremstyle{definition}

\theoremstyle{remark}

\graphicspath{{figures/}{.}}

\begin{document}
\articletype{AUTHOR'S ORIGIAL MANUSCRIPT (preprint version)}



\title{Effects of stoichiometry on premixed flames propagating in narrow channels: symmetry-breaking bifurcations}



\author{
  \name{Daniel Fern\'andez-Galisteo$^{\rm a}$\thanks{This is an Author's Original Manuscript (preprint) of an article published by Taylor $\&$ Francis Group in Combustion Theory and Modelling on 23/06/2017, available online: \newline \url{http://www.tandfonline.com/doi/full/10.1080/13647830.2017.1333157}}
    , Carmen Jim\'enez$^{\rm a}$, Mario S\'anchez-Sanz$^{\rm b}$ and Vadim N. Kurdyumov$^{\rm a}$} 
\affil{$^{\rm a}$Department of Energy, CIEMAT. Avda. Complutense 40, 28040 Madrid, Spain; $^{\rm b}$Dept. Ingenier\'ia T\'ermica y de Fluidos, Universidad Carlos III de Madrid, 28911, Legan\'es, Spain}}

\maketitle

\begin{abstract}

Recent studies within the diffusive-thermal (constant-density) approximation have shown that, for premixed flames freely propagating in narrow adiabatic channels, the instabilities induced by differential diffusion may result in non-symmetric solutions and/or oscillating and rotating propagation modes. This has been shown in the context of lean mixtures, for which a single species transport equation with a single Lewis number (corresponding to the ratio of thermal to molecular diffusivity of the deficient reactant) can be used to describe the flame propagation problem. In the present work we extend the analysis to mixtures of any equivalence ratio. To this end, we consider a two-reactant model, where the different diffusivities of the two reactants introduce two different Lewis numbers.
Steady-state computations and linear stability analysis are carried out for mixtures with large disparity between the Lewis number of the fuel ($Le_F$) and the oxidizer ($Le_O$), such as hydrogen-oxygen systems. It is shown that both differential diffusion and preferential diffusion have influence on the stability of the symmetric flame shape. For sufficiently lean and rich mixtures, the flame behaves as dictated by the Lewis number of the deficient reactant, i.e., the flame destabilizes toward non-symmetric solutions for large mass flow rates when the mentioned Lewis number is less than one. In near-stoichiometric mixtures the stability of the symmetric flame depends on a weighted average value of $Le_F$ and $Le_O$. In particular, the symmetric solution is stable for large mass flow rates because of the difficulties found by the less diffusive reactant to reach the reactive zone of the flame.


\end{abstract}





\section{Introduction}

Apart from the traditional safety implications concerning the possibility of flame ignition and propagation along a duct filled with a fuel and oxidizer mixture, the problem of flame propagation in narrow channels has received renewed attention in the last decade due to its role in new technologies for microflow reactors. Understanding the relevant flame structure and dynamics is fundamental for the future development of micro-combustors for power generation and propulsion systems \cite{Pe02,DuLeWa05,JuMa11} and the knowledge gained during last years has contributed to recent patented inventions \cite{JeMi09}.

Different mechanisms play a role in determining the structure and dynamics of the problem of a flame propagating in a narrow channel. We can mention, for example, the flame-fluid interaction produced by thermal expansion, the flame-wall heat exchange, the diffusive-thermal effect or the chemical complexities. The interplay of these physical phenomena acting simultaneously can obscure the role that each mechanism plays on the problem so it results convenient to study these effects separately, although perhaps some simplifications may create difficulties when comparing with experimental results. This is the line followed in preceding studies, where the effect of thermal expansion \cite{ShKe09,PeDa14,KuMa15}, heat losses to the walls \cite{DaMa02,SaFeKu14,KuJi14}, detailed chemistry and transport \cite{PiFrMaToBo08,PiMaFrToBo09,JiFeKu15} or differential diffusion \cite{DaMa01,KuFe02,Ku11,FeJiSaKu14} was investigated separately. The first asymptotic studies on flame propagation in narrow channels (or thick flames) already included the effect of heat losses, differential diffusion and the influence of the flow \cite{DaDoMa02}.

In particular, instabilities induced by the differential diffusion effect have been shown to result in non-symmetric flame solutions and/or oscillating and rotating propagation modes for premixed flames freely propagating in narrow adiabatic channels \cite{Ku11}. To isolate the effect of differential diffusion, these studies were conducted within the well-known diffusive-thermal approximation and in the context of a lean mixture, where a deficient reactant and a single Lewis number (characterizing the ratio between thermal and molecular diffusivities) were sufficient to describe the problem. It was demonstrated that steady non-symmetric solutions appear for large mass flow rates when the Lewis number is smaller than one and oscillating and pulsating flames for Lewis numbers significantly larger than one (and typically larger than 4 \cite{Ku11,FeJiSaKu14}). Here, we revisit this problem and investigate whether diffusive-thermal instabilities will also lead to non-symmetric solutions for flames with different stoichiometry and whether the global stability criterion proposed earlier for lean flames \cite{Ku11} needs to be modified.

To this end, in the present work we retain the diffusive-thermal approximation, which has been shown to be an excellent assumption for the study of a flame confined in a channel of width tens of the flame thickness or less and where only small-scale instabilities associated with the differential diffusion can emerge. We adopt a two-reactant model, where the different diffusivities of the two reactants introduce two different Lewis numbers, and consider the case of a large disparity between these two Lewis numbers, as typically found in hydrogen-oxygen systems. It will be shown that the global stability will depend on the relative values of these two Lewis numbers as well as on the equivalence ratio, a concept earlier introduced by Joulin and Mitani \cite{JoMi81} through an effective Lewis number of the mixture.


\section{General formulation}

\begin{figure}[ht]
\begin{center}
  \includegraphics[width=0.9\textwidth]{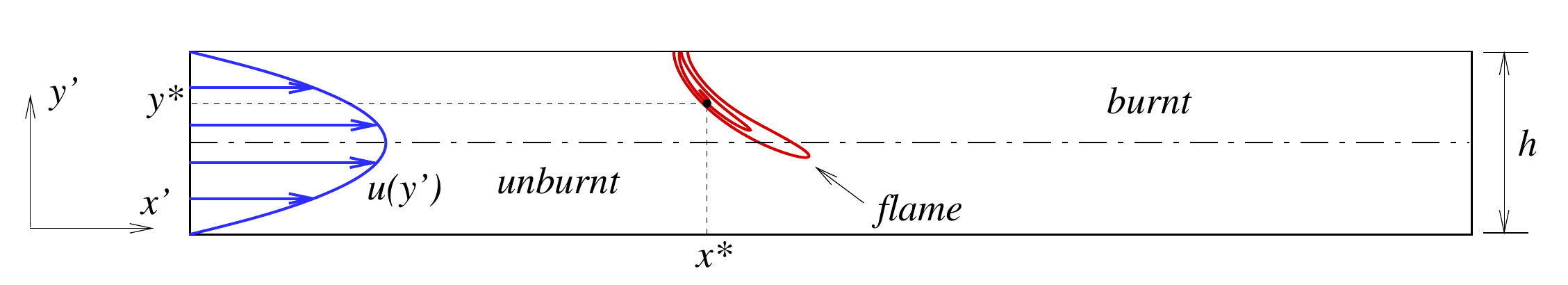}
\caption{Sketch of the problem, coordinate system, parabolic velocity field and the typical reaction rate isocontour of a steady non-symmetric flame shape. Also shown is an example of the location of the reference point ($x^*,y^*$), which is used to attach the flame and calculate the propagation velocity from Eqs.~\eqref{eq:T}-\eqref{eq:Y} written in a moving reference frame.}
\label{fig:sketch}
\end{center}
\end{figure}

Consider a premixed flame propagating with velocity $U_f$ in a planar adiabatic channel of height $h$, as sketched in Fig.~\ref{fig:sketch}. The fuel-air mixture is at initial temperature $T_u$ and driven by a Poiseuille flow with mean velocity $U_0$. In the diffusive-thermal approximation, the mixture density $\rho$, the mixture heat capacity $c_p$, the mixture thermal diffusivity $\mathcal{D}_T$, and the molecular diffusivity of the reactants $\mathcal{D}_i$ are all assumed to be constant, where $i$ stands for the species index. As a consequence, the flow field is not affected by combustion and we concentrate on instabilities induced by the differential diffusion. The inflow velocity is given by the Poiseuille's law $u(y)=6U_0(y/h)(1-y/h)$.

The chemical reaction is modeled through a one-step irreversible reaction $\nu_F F + \nu_O O\rightarrow {\rm Products}\, + \,Q$, where $F$ and $O$ denote the chemical symbols for fuel and oxidizer, $\nu_F$ and $\nu_O$ are the corresponding molar stoichiometric coefficients, and $Q$ is the heat of combustion per $\nu_F$ mole of fuel consumed. The reaction is supposed to take place in an inert diluent species (i.e., nitrogen when air is utilized). The molar reaction rate is assumed to have an Arrhenius form, with first order with respect to each reactant, $\omega=\mathcal{B}\rho^2(W_F W_O)^{-1} Y_F Y_O \exp{(-E/\mathcal{R}T)}$, where $\mathcal{B}$ is the frequency factor, $E$ the global activation energy, $\mathcal{R}$ the universal gas constant, $T$ the temperature of the mixture, $Y_F$ and $Y_O$  the local mass fraction of the fuel and oxidizer respectively, and $W_F$ and $W_O$ the corresponding molecular weights. 


The equivalence ratio of the mixture is defined as $\phi=s Y_{F_u}/Y_{O_u}$, where $s=\nu_O W_O/(\nu_F W_F)$ is the mass stoichiometric ratio and measures the mass of oxidizer needed to burn the unit of mass of fuel, and $Y_{F_u}$ and  $Y_{O_u}$ correspond to the mass fraction of fuel and oxidizer in the fresh unburnt region at $x\rightarrow -\infty$. To avoid the discussion of lean and rich mixtures separately, a convenient parameter $\Phi=(\nu_1 W_1 Y_{2_u})/(\nu_2 W_2 Y_{1_u})$, which measures the ratio of mass of abundant-to-deficient reactants in the fresh mixture, is introduced, as done before in \cite{BeMa99}. The subscripts $1$ and $2$ stand for the deficient and abundant reactants, respectively, and therefore replace the subscripts $F$ and $O$ as appropriate, except for a stoichiometric mixture ($\Phi=1$) when the terms ``deficient'' and ``abundant'' do not describe the mixture, in the sense that no reactant is present in the burnt gases. Otherwise, the parameter is always larger than one and corresponds with the equivalence ratio, $\Phi=\phi$, for fuel-rich mixtures and to its inverse, $\Phi=\phi^{-1}$, for fuel-lean mixtures. 

\subsection{Dimensionless governing equations}

The channel width $h$ and the diffusion time $h^2/\mathcal{D}_T$ are chosen as the reference units of length and time to define the non-dimensional coordinates $x=x^{\prime}/h$, $y=y^{\prime}/h$ and the time $t=t^{\prime}/(h^2/\mathcal{D}_T)$. The non-dimensional temperature is defined as $\theta=(T-T_u)/(T_a-T_u)$, with $T_a=T_u+Q Y_{1_u}/(c_p\, \nu_1 W_1)$ the adiabatic flame temperature. The fuel and oxidizer mass fractions are normalized with the fresh values in the form $Y_1=Y_1^{\prime}/Y_{1_u}$ and $Y_2=\Phi Y_2^{\prime}/Y_{2_u}$, respectively. Primes indicate dimensional or non-reduced quantities here and hereafter.

Introducing the burning velocity of the planar flame, $S_L$, and the thermal flame thickness, $\delta_T=\mathcal{D}_T/S_L$, in a coordinate system attached to the flame, $x \rightarrow x - u_f t$, where $u_f=U_f/S_L$ is the reduced propagation velocity of the flame with respect to the wall, the dimensionless governing equations are identical to those presented before in \cite{SaFeKu14}, in the context of a different problem:
\begin{align}
\p{\theta}{t}+\sqrt{d}\{u_f+6\,my(1-y)\} \p{\theta}{x}&=\Delta \theta + d\,\omega, \label{eq:T}\\ 
\p{Y_i}{t}+\sqrt{d}\{u_f+6\,my(1-y)\} \p{Y_i}{x}&=\frac{1}{Le_i}\Delta Y_i - d\,\omega \qquad i=1,2, 
\label{eq:Y}
\end{align}
where the reaction rate is given by 
\begin{equation}
\omega = \frac{\beta^2}{2 \mathcal{L} s_L^2}Y_1Y_2\exp{\left\{\frac{\beta(\theta-1)}{1+\gamma(\theta-1)}\right\}},
\label{eq:w}
\end{equation}
with 
\begin{equation*}
\mathcal{L}=Le_1Le_2(1+\mathcal{A})/\beta \qquad \text{and} \qquad \mathcal{A}=1+ \beta(\Phi-1)/Le_2.
\end{equation*}

The following parameters appear in the formulation: the relevant Zel'dovich number $\beta=E(T_a-T_u)/\mathcal{R}T_a^2$, the heat release parameter $\gamma=(T_a-T_u)/T_a$, the Lewis number of the deficient and abundant reactants $Le_1$ and $Le_2$, respectively, the reduced mass flow rate $m=U_0/S_L$, and the Damk$\ddot{\rm o}$hler number $d=(h/\delta_T)^2$. 
The symbol $\Delta=\partial^2/\partial {x}^2 + \partial^2/\partial {y}^2$ stands for the Laplacian operator. 

The problem needs to be supplemented by the following boundary conditions far upstream and downstream of the flame front 
\begin{equation}
\begin{aligned}
  x\to -\infty:&\quad \theta=Y_1-1=Y_2-\Phi=0,  \\
x\to +\infty:&\quad \pl{\theta}{x}=\pl{Y_i}{x}=0, \qquad i=1,2.
\end{aligned}
\label{BC2Dx}
\end{equation}

At the walls we impose adiabatic and non-porous conditions
\begin{equation}\label{BC2Dy}
\begin{aligned}
y=0 \quad \text{and} \quad y=1: \quad \pl{\theta}{y}=\pl{Y_i}{y}=0 \qquad i=1,2.
\end{aligned}
\end{equation}

Anticipating the numerical results, two kind of solutions can emerge during the steady-state computations: symmetric and non-symmetric. When the flame results non-symmetric, we force the symmetric solution separately by reducing the domain to half its height and imposing symmetric boundary conditions $\pl{\theta}{y}=\pl{Y_1}{y}=\pl{Y_2}{y}=0$ at $y=1/2$. The symmetric solution is used in the linear stability analysis of Section \ref{sec:stability}.

The equilibrium value of the fuel and oxidizer mass fractions sufficiently far downstream the flame can be obtained by integrating a combination of the species transport equation \eqref{eq:Y} to give $Y_1=0$ and $Y_2=\Phi-1$ at $x\to +\infty$.

Note that the above formulation can be simplified in the limit of lean and rich mixtures, when the problem reduces to solving the equations of energy and mass fraction of the deficient reactant. In particular, for $\Phi \rightarrow \infty$, the problem simplifies to that presented before in \cite{Ku11}, with the mass fraction of the abundant reactant remaining constant ($Y_2=1$) and with $\mathcal{L}= Le_1$.

The factor $s_L=S_L/(S_L)_{asp}$ introduced in (\ref{eq:w}), corresponds to the eigenvalue of the planar adiabatic problem
\begin{equation}
\begin{aligned}
\der{\theta}{\xi}&=\dd{\theta}{\xi} + \omega,  \\ 
\der{Y_i}{\xi}&=\frac{1}{Le_i}\dd{Y_i}{\xi} - \omega, \qquad i=1,2,\\
\label{eq:planar}
\end{aligned}
\end{equation}
where $\xi=x\sqrt{d}$, to be solved with the boundary conditions given in (\ref{BC2Dx}). The factor $s_L$ ensures that, for a given value of $\Phi$, the burning velocity of the flame propagating in the channel, $u_f$, equals one when the flame acquires the planar shape. This factor represents the ratio between the planar flame speed obtained with finite activation energy, $S_L$, and the flame speed calculated by assuming infinite activation energy \cite{SeLu79,BeMa99} 
\begin{equation}
(S_L)_{asp}^2 = 2\mathcal{B} \rho\mathcal{D}_T Y_{1_u} \frac{\nu_2 Le_1 Le_2}{W_1\beta^3} \left(1+\mathcal{A}\right) e^{-E/\mathcal{R}T_a}.
  \label{eq:SLasp}
\end{equation}
The use of the subindex $i$ and the parameter $\Phi$, as presented above, results in a simplified unique formulation, which is very practical for the implementation of the numerical simulations. During the presentation of the results it is more convenient to distinguish between lean and rich mixtures so, in what follows, the subscripts notation $F$ and $O$ is recovered again together with the definition of the equivalence ratio $\phi$.

\section{Global flame parameters}
Previous theoretical analyses using the two-reactant combustion model \cite{SeLu79,Ja87,JoMi81} have not considered the dependence of the adiabatic temperature, the activation energy or the Lewis numbers of the reactants with the stoichiometric ratio. This was partially justified because the analysis was carried out at near-stoichiometric conditions, although the assumption can also be justified in some cases for off-stoichiometric mixtures. For example, Shen et al.~\cite{ShWoGrMaRo16}, in their experiments on hydrogen-oxygen flames, maintained a constant adiabatic temperature while varying the equivalence ratio by adequately diluting the mixture with a variable amount of inert gas. 
The conservation of mass indicates that $Y_{F_u}+Y_{O_u}=1-Y_{N_u}=c$, where $Y_{N_u}$ stands for the mass fraction of the inert gas in the fresh unburnt mixture; see \cite{SeLu79}. Changing $Y_{N_u}$ in an adequate proportion (or alternatively the parameter $c$) the adiabatic temperature can be held constant in the experiments.

Following this approach, we have calculated in Table~\ref{table:properties} the dilution parameter $c$ required to maintain the adiabatic temperature constant and equal to $T_a=1600$ K ($\gamma=0.8$) for lean, stoichiometric and rich flames of hydrogen, methane, methanol and propane diluted with nitrogen at $p=1$ atm and $T_u=300$ K. Additionally, we show in the table the flame speed $S_L$ (computed using the recently revised San Diego mechanism \cite{SD16}), the activation energy $E$ and its associated Zel'dovich number $\beta$, calculated as described in \cite{EgLa90b} by the expression $E=-2\mathcal{R}\left\{ \pl{[\ln{(\rho_u S_L)]}}{(1/T_a)}\right\}$ with use of the flame speed $S_L$ and the adiabatic flame temperature $T_a$. The parameter $\beta$ falls within the well-known values $5 \leqslant \beta \leqslant 15$ \cite{Wi85} and shows values independent of the composition for moderately off-stoichiometric flames ($0.6\leqslant \phi \leqslant 2$) with deviations less than 10$\,\%$, at least for the dilution degrees utilized. The differences become much larger outside this range, reaching a 21$\,\%$ deviation for hydrogen flames at extremely rich conditions ($\phi=5$) or for the propane case. With this in mind, we decide to consider in the computations constant values of $\beta=10$ and $\gamma=0.8$.

The constant Lewis number approximation has been used extensively in previous theoretical studies of the stability of two-reactants flames \cite{JoMi81,Ec16} but its applicability is compromised when highly diffusive fuels are used. In Table~\ref{table:properties} we show the Lewis numbers of fuel and oxidizer estimated in the fresh region of planar flames computed using a multicomponent diffusion model. From the calculated values of $Le_F$ and $Le_O$, we check that they remain fairly constant with the equivalence ratio for all fuels but hydrogen. Results of the single-reactant model \cite{Ku11} indicate that only cases with Lewis number less than unity exhibit non-symmetric solutions for a flame subject to a Poiseuille flow. In the two-reactant model, the Lewis number of the limiting reactant determines the stability behavior of the flame when the other reactant (oxidizer in the case of lean mixtures or fuel in the case of rich ones) is in excess \cite{Wi85}. In this sense, the selection of $Le_F=0.3$ for the Lewis number of the fuel (representative of lean hydrogen-oxygen mixtures) and $Le_O=2$ for the Lewis number of the oxidizer (representative of rich hydrogen-oxygen mixtures \cite{SuSuHeLa99}) best illustrates a complete range of flame behaviors and for this reason is considered in the present paper. 
For completeness, we also consider the case with $Le_O=1$.

\begin{ctable}[
caption={Global flame parameters for lean, stoichiometric and fuel-rich mixtures diluted in nitrogen at $p=1$ atm and $T_u=300$ K. For all cases the dilution factor $c$ is such that $T_a=1600$ K ($\gamma=0.8$).}, 
label={table:properties},pos=h]
{c c c c c c c c c}{}
{\toprule
  Fuel & $\phi$ & $c$ & $S_L$ & $E$ & $\beta$ & $Le_F$ & $Le_O$   \\
       &        &     & [cm/s]  & [kcal/mol] &  &          \\
\midrule
H$_2$    & 0.5 & 0.23 & 46.4    & 32.8   & 8.4       & 0.35 & 1.46  \\
H$_2$    & 1.0 & 0.12 & 40.4    & 31.7   & 8.1       & 0.35 & 1.43  \\
H$_2$    & 5.0 & 0.56 & 207.3   & 24.9   & 6.4       & 0.35 & 2.59   \\
\midrule
CH$_4$   & 0.6 & 0.24 & 8.1     & 48.1   & 12.2      & 0.98 & 1.10  \\
CH$_4$   & 1.0 & 0.16 & 3.4     & 55.4   & 14.2      & 0.98 & 1.08  \\
CH$_4$   & 1.3 & 0.19 & 2.3     & 51.8   & 13.2      & 0.96 & 1.09  \\
\midrule
CH$_3$OH & 0.6 & 0.26 & 7.9     & 42.9   & 11.0      & 1.38 & 1.05  \\
CH$_3$OH & 1.0 & 0.19 & 7.7     & 39.7   & 10.1      & 1.37 & 1.04  \\
CH$_3$OH & 2.0 & 0.37 & 9.0     & 35.0   & 9.1       & 1.24 & 0.98   \\
\midrule
C$_3$H$_8$  & 0.6 & 0.24 & 9.9  & 42.9   & 10.9      & 1.85 & 1.06 \\
C$_3$H$_8$  & 1.0 & 0.16 & 4.7  & 51.3   & 13.1      & 1.85 & 1.05 \\
C$_3$H$_8$  & 1.6 & 0.23 & 2.4  & 58.4   & 14.9      & 1.80 & 1.02 \\

\bottomrule}
\end{ctable}

\section{Stationary symmetric and non-symmetric flames}\label{sec:steady}

We integrate the steady version of the system of equations \eqref{eq:T}-\eqref{eq:Y}, with boundary conditions \eqref{BC2Dx}-\eqref{BC2Dy}, using a Gauss-Seidel iterative method with over-relaxation to obtain the temperature and mass fraction fields together with the propagation velocity $u_f$ for given values of the Damkh$\ddot{\rm o}$ler number $d$, the Lewis number of the deficient $Le_1$ and abundant reactant $Le_2$, and the mass flow rate $m$. The spatial derivatives are discretized using second order, three-point central finite differences on a rectangular grid with a resolution of at least 10 grid points in the reaction zone. The invariability of the equations under translations in the $x$-coordinate adds an extra condition to obtain the eigenvalue $u_f$ by imposing a value for the temperature $T=T^*$ at some point $(x,y)=(x^*,y^*)$, with $T^*=0.7$ and $(x^*,y^*)=(0,0.8)$ typical values used during the computations.\\
We characterize the solution by calculating the eigenvalue $u_f$ and the symmetry of the flame by means of the parameter $S$, defined as
\begin{align}
S=\int_{-\infty}^{+\infty} \int_0^{1/2} {\rm abs}\{\theta(x,y)-\theta(x,1-y)\} \, {\rm d}y \,{\rm d}x,
\end{align}
where abs$\{\cdot\}$ stands for the absolute value. By definition, the flame remains symmetric when $S=0$ and becomes asymmetric when $S>0$. Obviously, each non-symmetric solution has an associated mirror counterpart. The accuracy of the numerical solution is checked by comparing the results computed with our code in the limit $\phi \to 0$ with the results previously calculated by Kurdyumov in \cite{Ku11}, giving a satisfactory match. 

\begin{figure}[!ht]
\begin{center}
  \includegraphics[width=0.49\textwidth]{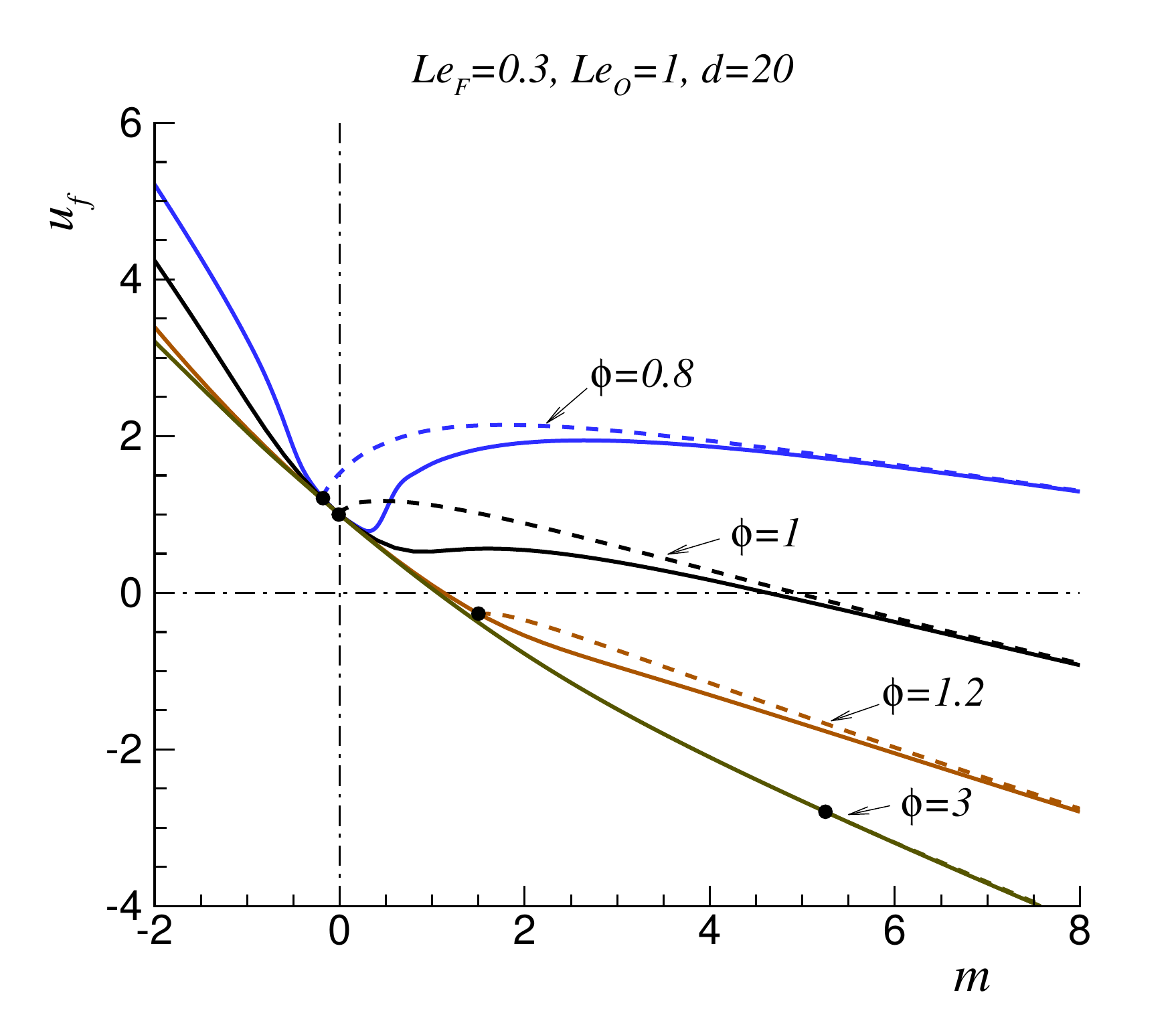}
  \includegraphics[width=0.49\textwidth]{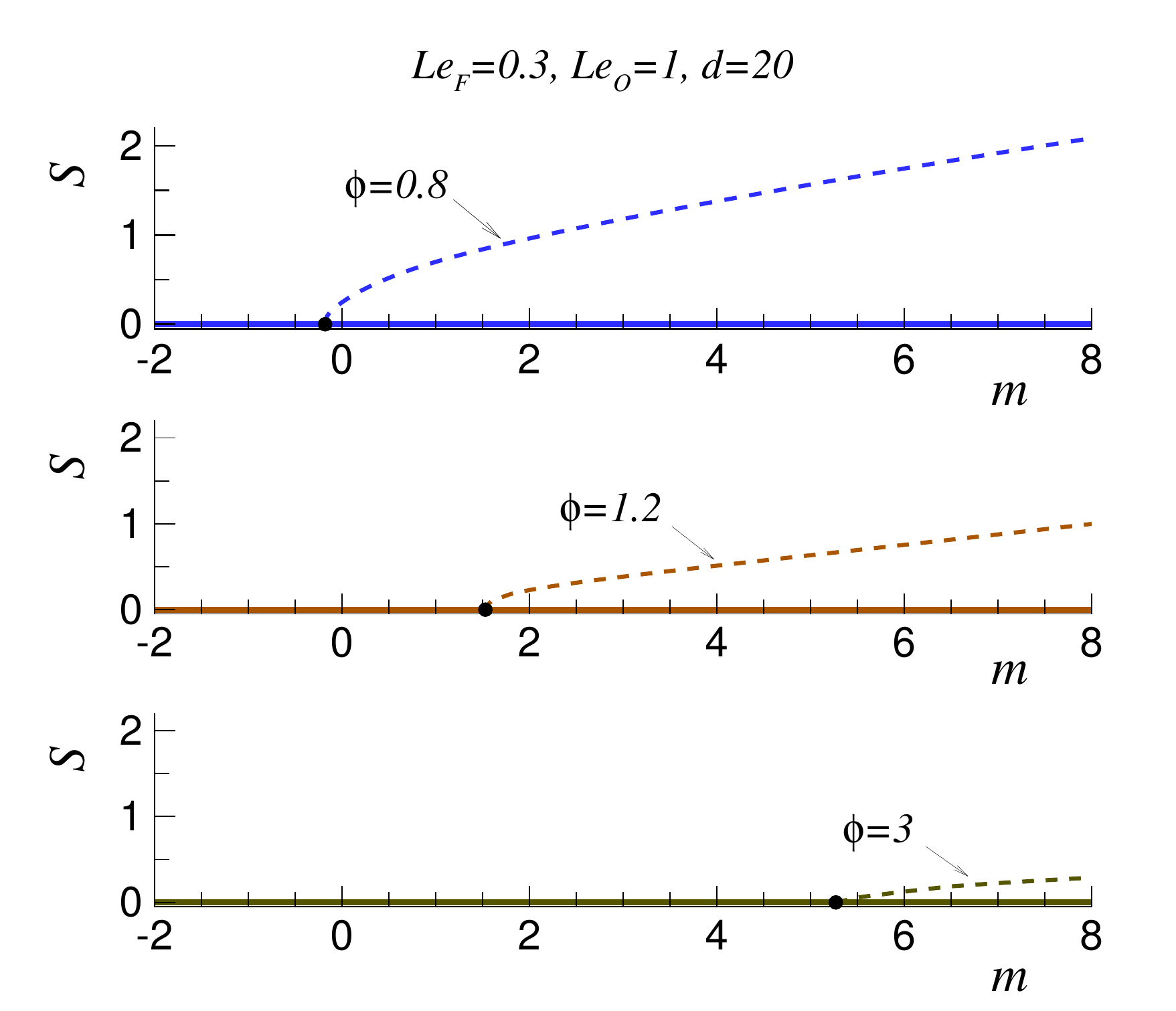}    
\caption{The variation with the flow rate of the propagation velocity $u_f$ (left) and the value of the parameter $S$ (right). Calculated for $Le_F=0.3$, $Le_O=1$ and $d=20$.}
\label{fig:uf_m_d20_LeO-1}
\end{center}
\end{figure}

\begin{figure}[!ht]
\begin{center}
  \includegraphics[width=0.49\textwidth]{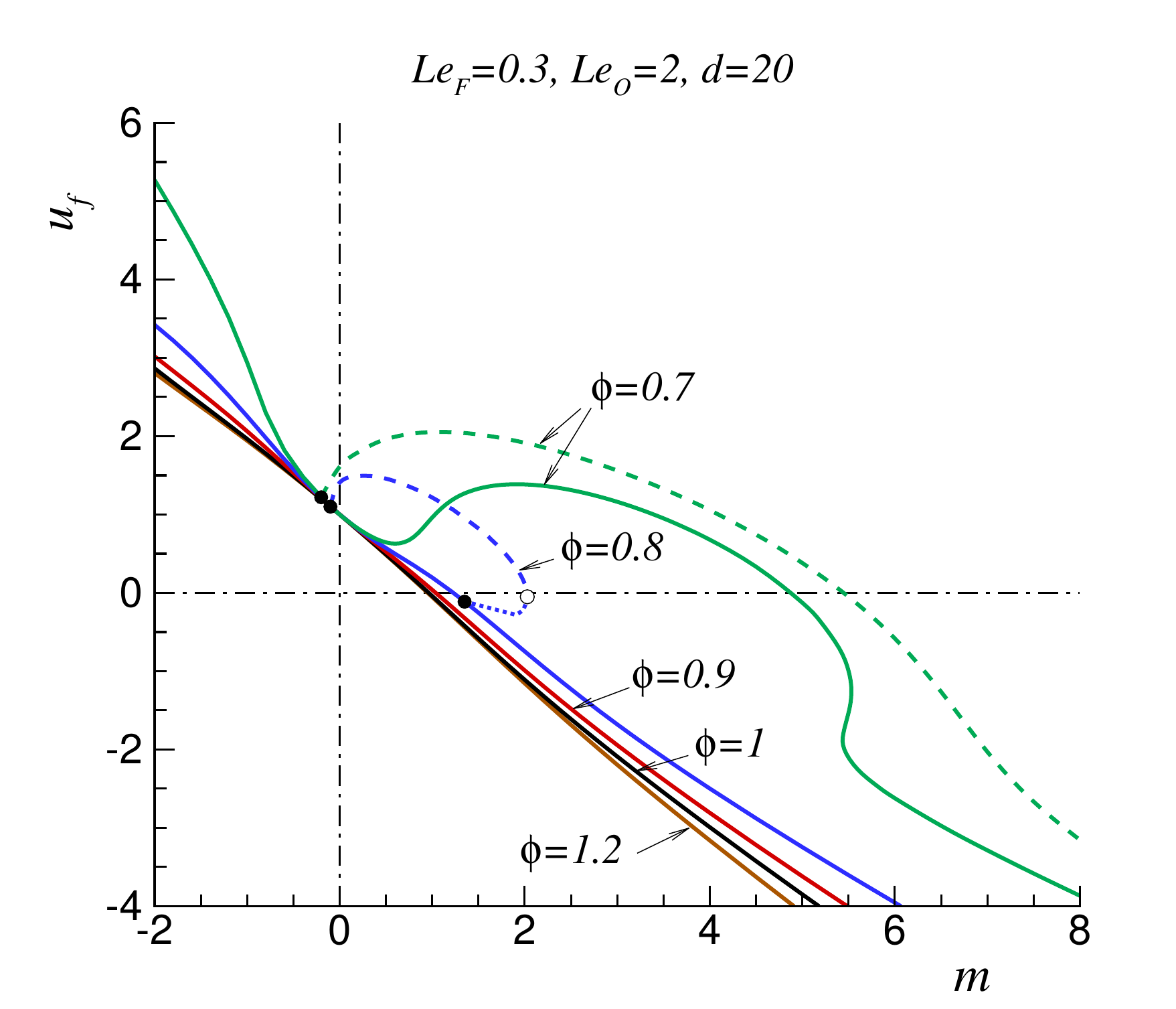}
   \includegraphics[width=0.49\textwidth]{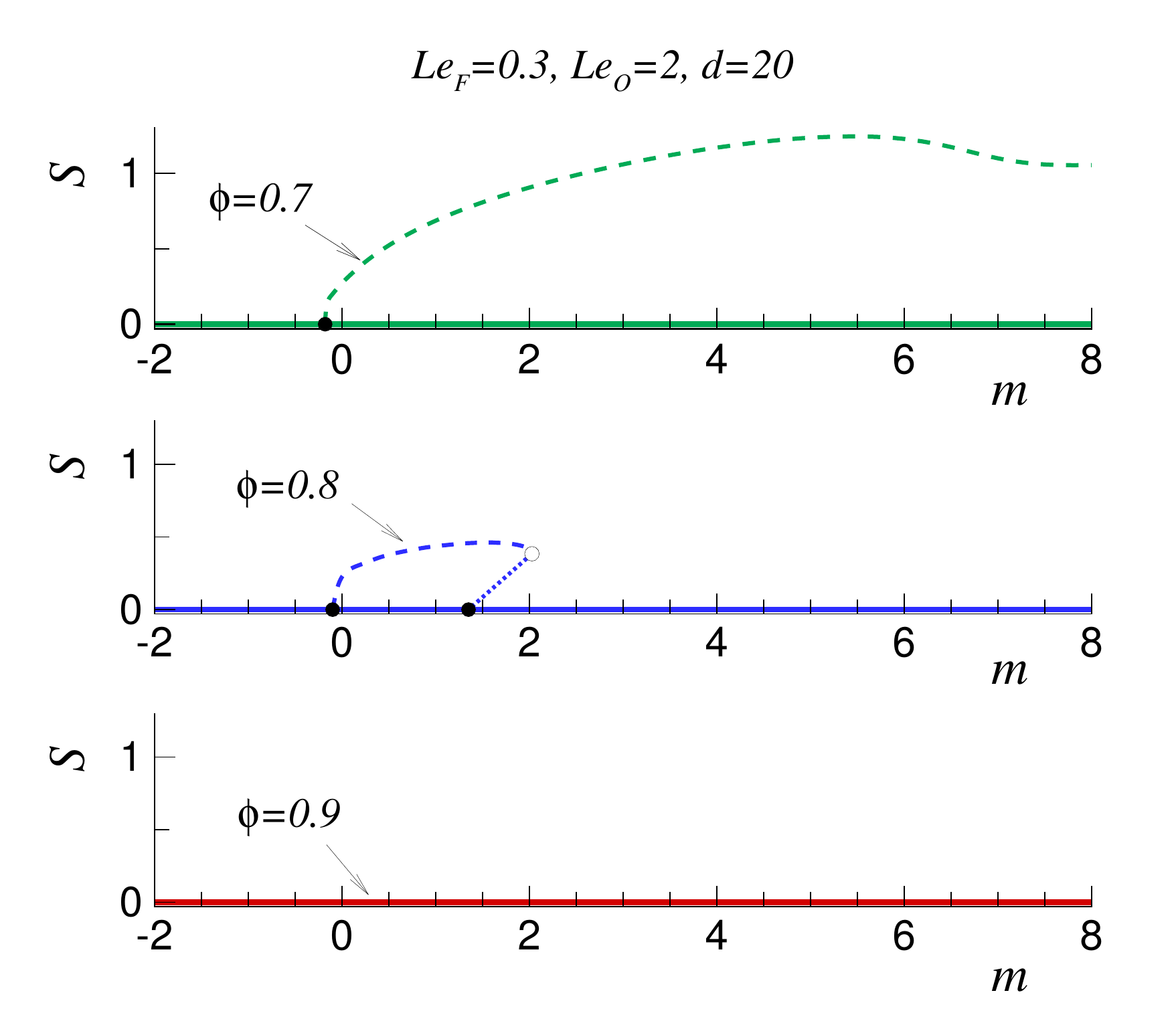}
\caption{The variation with the flow rate of the propagation velocity $u_f$ (left) and the value of the parameter $S$ (right). Calculated for $Le_F=0.3$, $Le_O=2$ and $d=20$. The dotted line represents an unstable non-symmetric branch, as calculated in the next section. Calculated for $Le_F=0.3$, $Le_O=2$ and $d=20$. 
}
\label{fig:uf_m_d20}
\end{center}
\end{figure}

\begin{figure}[!ht]
\begin{center}
  \includegraphics[width=0.49\textwidth]{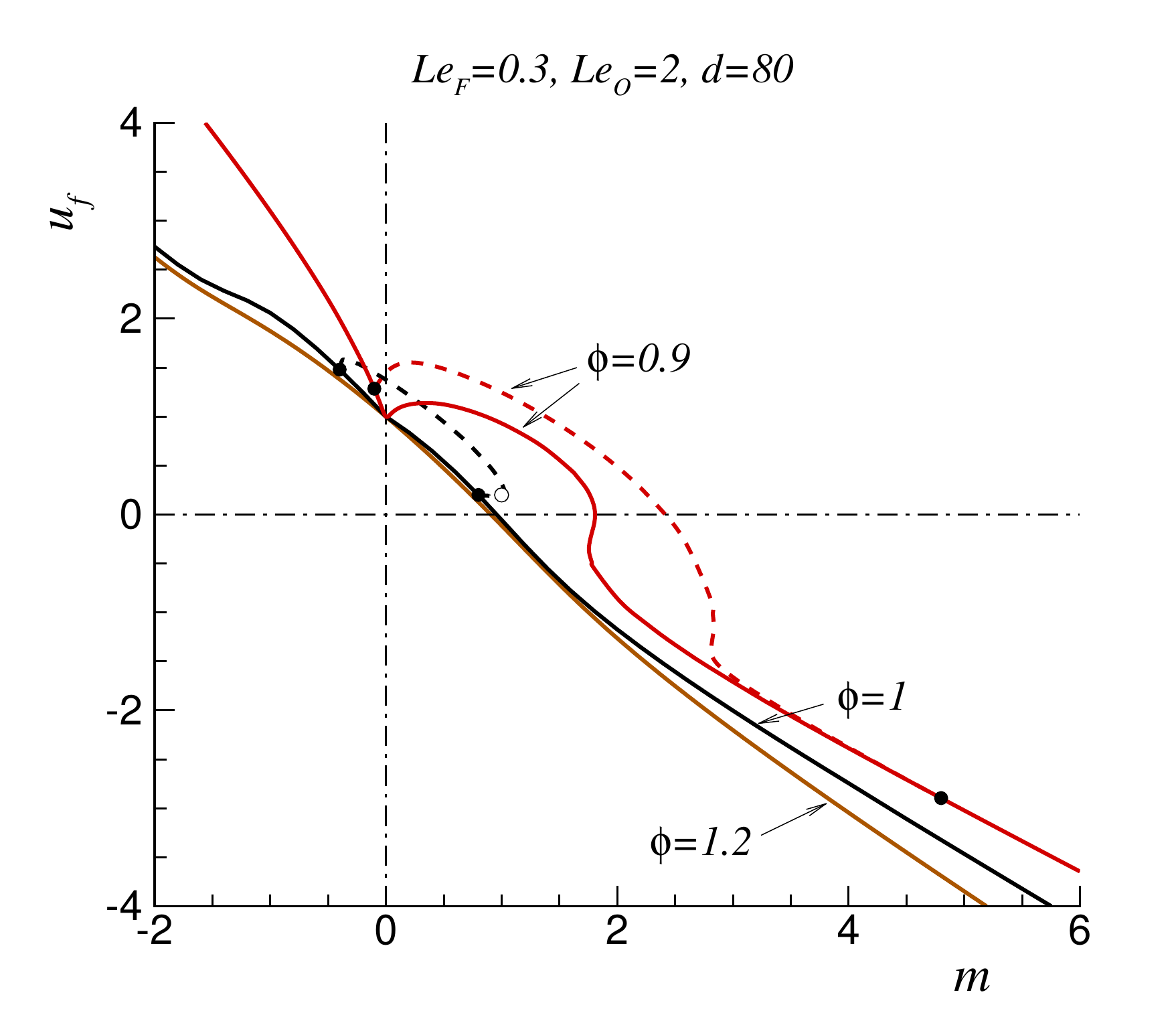}
  \includegraphics[width=0.49\textwidth]{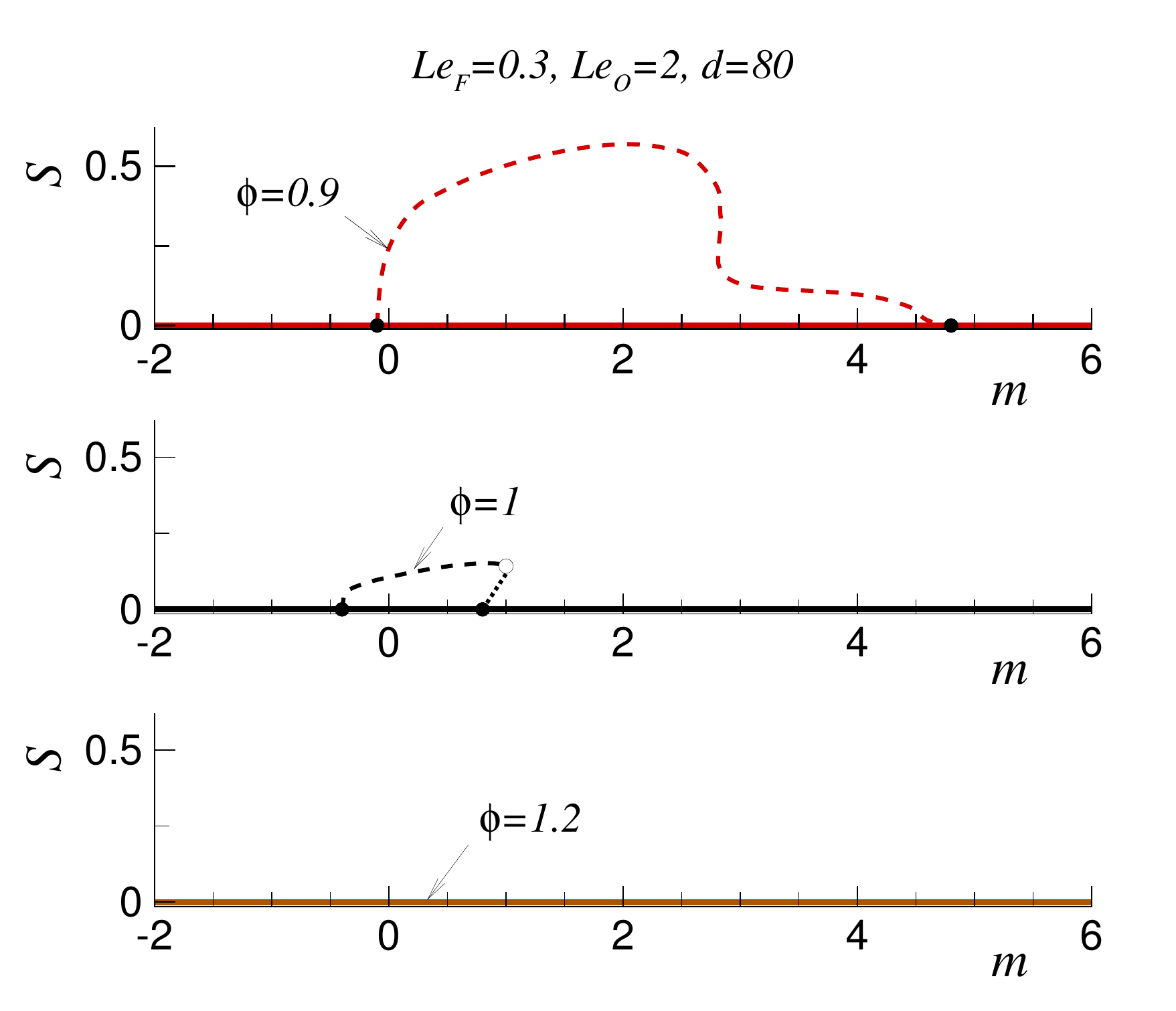}
\caption{The variation with the flow rate of the propagation velocity $u_f$ (left) and the value of the parameter $S$ (right). Calculated for $Le_F=0.3$, $Le_O=2$ and $d=80$.}
\label{fig:uf_m_d80}
\end{center}
\end{figure}

The calculated values of the eigenvalue $u_f$ versus the mass flow rate $m$  are plotted in Figs.~\ref{fig:uf_m_d20_LeO-1}, \ref{fig:uf_m_d20} and \ref{fig:uf_m_d80} for $Le_F=0.3$ and  several equivalence ratios $\phi$, oxidizer diffusivity $Le_O$ and Damk$\ddot{\rm o}$hler numbers $d$. Solid curves represent symmetric solutions and dashed curves represent non-symmetric solutions. 

Let us first address the case $Le_F=0.3$, $Le_O=1$ in a narrow channel $d=20$. As depicted in Fig.~\ref{fig:uf_m_d20_LeO-1} (left), when the mass flow rate parameter $m$ is increased from negative (assisted flow) towards positive values (opposed flow) we find a (supercritical) bifurcation point at $m=m_{b_1}$ (marked with $\bullet$), different for each equivalence ratio. The linear stability analysis of the symmetric solution, presented in next section, indicates that the symmetric solution becomes unstable at this point and the flame acquires a non-symmetric shape. For large mass flow rates the propagation velocity of the non-symmetric solutions tends to be equal to that of the corresponding symmetric flame, because the structure of both solutions in the near-wall region are similar and the propagation velocity is mainly determined by the flow velocity gradients in the wall \cite{KuFeLi00,Ku11}. The symmetry breaking is more clearly shown in Fig.~\ref{fig:uf_m_d20_LeO-1} (right) where the value of the parameter $S$ is plotted versus the flow rate. In this figure we see that $S$ becomes positive for $m>m_{b_1}$ when the non-symmetric solution emerges. For fuel-lean mixtures the bifurcation point tends asymptotically to a negative value close to zero ($m_{b_1} \to -0.21$) and the flame is non-symmetric for all positive values of $m$. For fuel-rich mixtures the bifurcation point moves to increasing values of $m$ and the non-symmetric solution only emerges for very large mass flow rates. 

The case $Le_O=2$ introduces a large disparity between the two mass diffusivities that brings forth double-valued regions and the re-stabilization of the symmetric solution for large values of $m$, features not observed for more diffusive oxidizers. See, for example, the second (subcritical) bifurcation point that emerges at $m=m_{b_2}=1.35$ (also marked with $\bullet$) for $\phi=0.8$ in Fig.~\ref{fig:uf_m_d20}. According to this figure, the flame acquires a symmetric shape for $m>1.35$ indicating that the Poiseuille flow stabilizes the symmetric flame, an effect not documented previously in single-reactant models \cite{Ku11}. Furthermore, symmetric and non-symmetric solutions coexist for $m_{b_2}<m<m_{b_3}$, delimiting a range of mass flow rates with multiplicity of solutions. The symbol $\circ$ in the non-symmetric branch indicates a fold bifurcation at $m=m_{b_3}=2.02$. The stability analysis, presented in next section, shows that part of this branch (marked with dotted line) is intrinsically unstable. For $\phi=0.7$ the second bifurcation point disappears and the non-symmetric shape is the only stable solution for $m>m_{b_1}$. It is worth mentioning that at this equivalence ratio one finds a region of multi-valued symmetric solutions for $5.45<m<5.55$ but all of these solutions are intrinsically unstable, as will be shown in the stability analysis presented below, so the non-symmetric shape is the only stable solution. As before, we recover the results of the single-reactant model for leaner mixtures. For fuel-rich mixtures the flame shape is always symmetric, as dictated by the stability criterion corresponding to a Lewis number larger than unity \cite{Ku11}.

\begin{figure}[ht]
\begin{center}
  \includegraphics[width=0.49\textwidth]{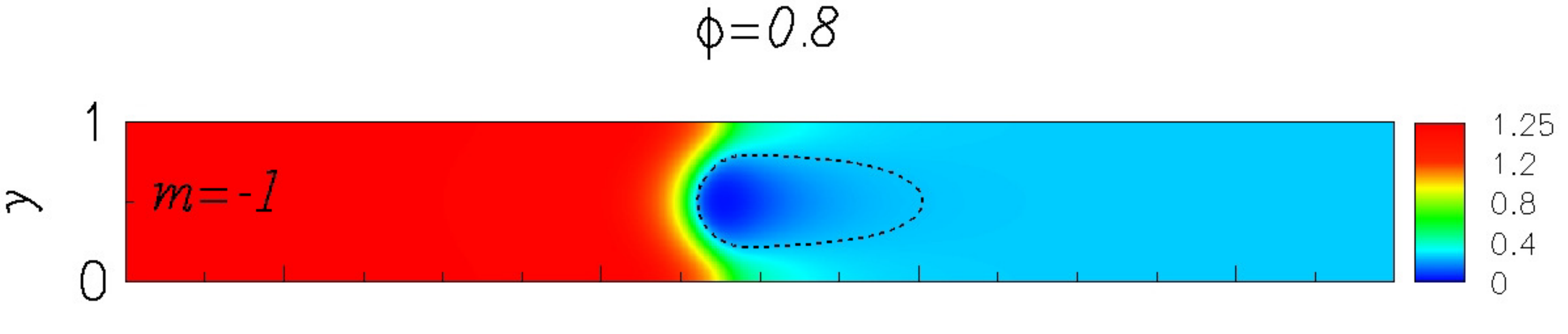}
  \includegraphics[width=0.49\textwidth]{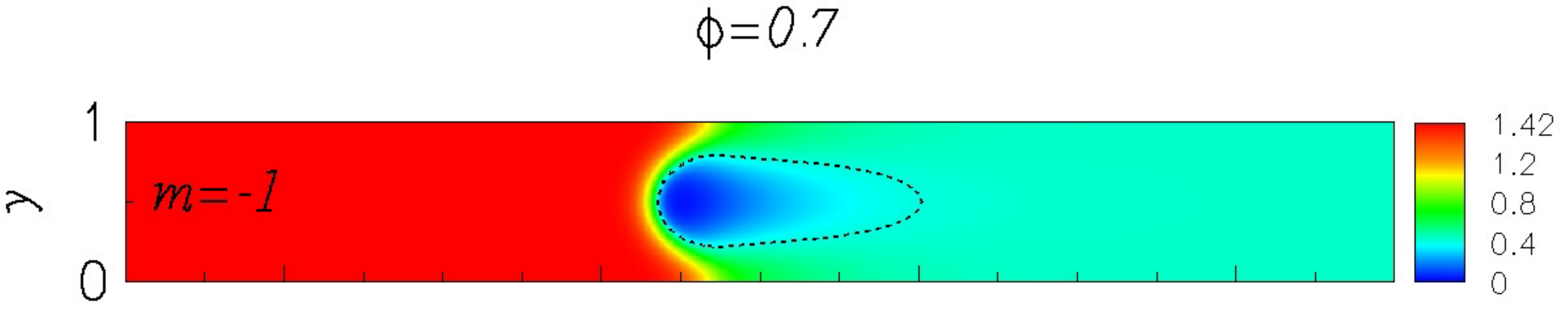} \\
  \includegraphics[width=0.49\textwidth]{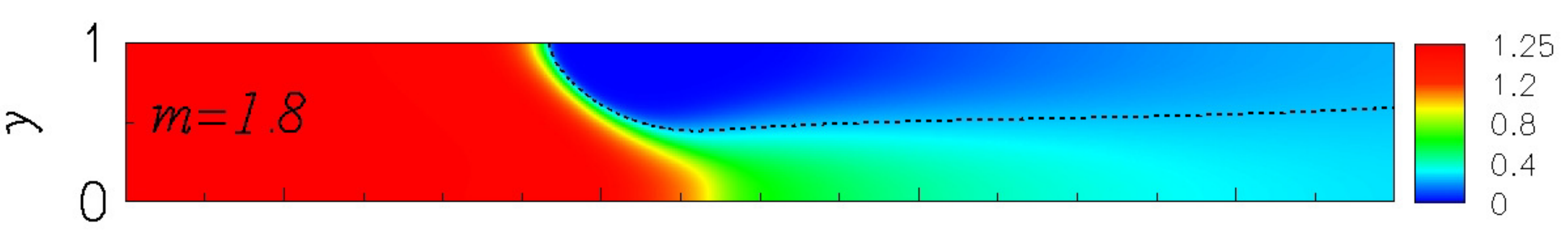}
  \includegraphics[width=0.49\textwidth]{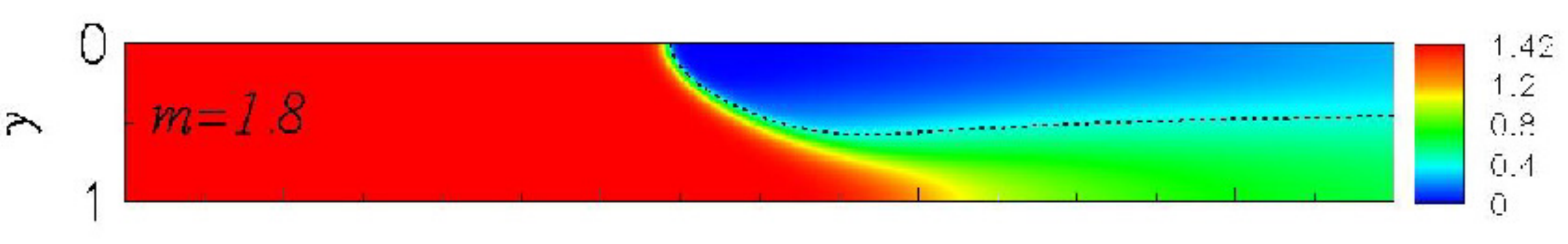} \\
  \includegraphics[width=0.49\textwidth]{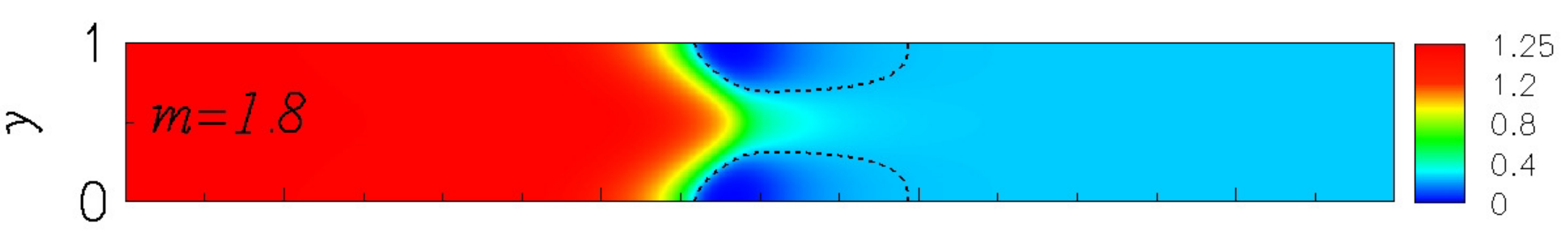}
  \includegraphics[width=0.49\textwidth]{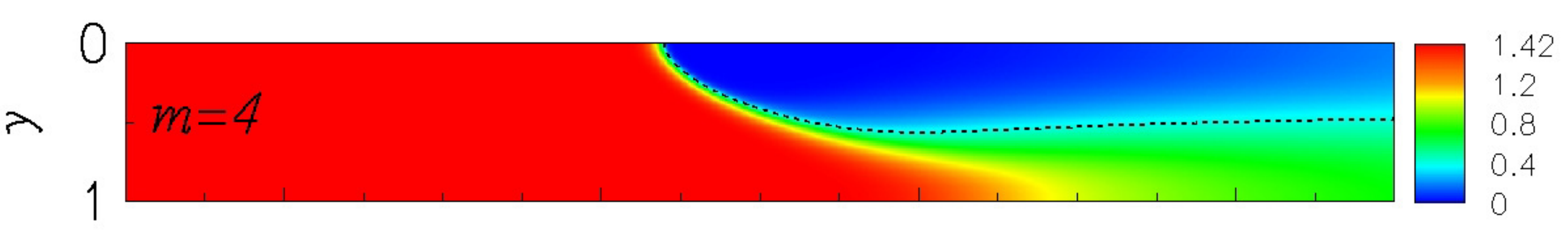}\\
  \includegraphics[width=0.49\textwidth]{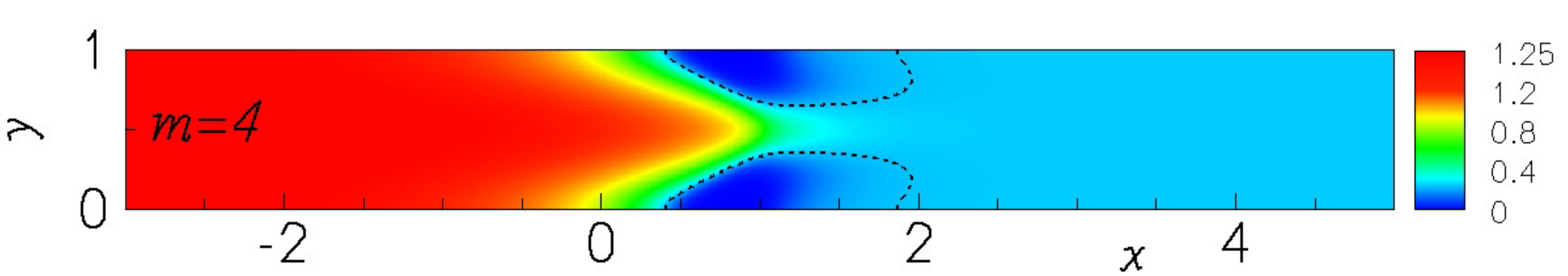}
  \includegraphics[width=0.49\textwidth]{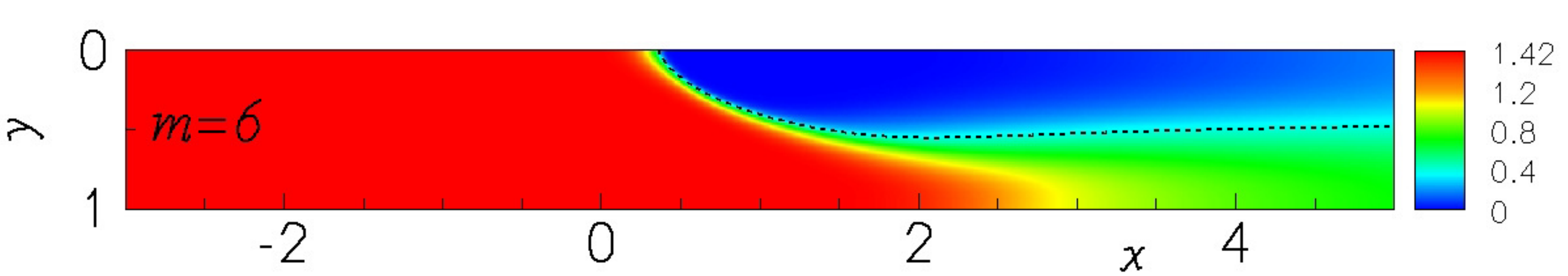}
\caption{(Color online) Structure of the flame front represented by the isocontour of the abundant mass fraction $Y_2$ for $\phi=0.8$ (left) and $\phi=0.7$ (right) at variable flow rates. Black dashed lines separate the region with mass fraction below the equilibrum value $Y_2(x\rightarrow \infty)=(\Phi-1)$. Calculated for $Le_F=0.3$, $Le_O=2$, and $d=20$.}
\label{fig:YO_d20}
\end{center}
\end{figure}

To get a better idea of the shape solutions, we plot in Fig.~\ref{fig:YO_d20} the flame structure given by the mass fraction isocontours of the abundant reactant $Y_2$ for the case $Le_O=2$. In the figure we compare the solutions with $\phi=0.8$ (left) and $\phi=0.7$ (right) for several mass flow rates. 
As seen before, with $\phi=0.8$ the solution admits both symmetric and non-symmetric shapes in a small range of flow rates and this is illustrated for $m=1.8$. Unlike the leaner flame $\phi=0.7$, the case with $\phi=0.8$ shows a stabilizing effect of the symmetric solution for large values of the mass flow rate and this is illustrated for $m=4$. Associated with the slow diffusion of the oxidant, we find a region near the flame (marked with dashed curve) within which the concentration of the species in excess falls below the value of equilibrium $(1-\phi)/\phi$. 

An increase of the channel width modifies the critical values where symmetry breaking occurs. This is related to the fact that the channel width determines the size of unstable modes than can grow inside the channel. Fig.~\ref{fig:uf_m_d80} presents the propagation velocity results obtained for a channel of width $d=80$, with $Le_F=0.3$ and $Le_O=2$. Non-symmetric solutions are found for $\phi=1$ and $\phi=0.9$ in a range $m_{b_1}<m<m_{b_2}$ ($m_{b_1}=-0.4$ and $m_{b_2}=0.8$ for $\phi=1$ and $m_{b_1}=-0.1$ and $m_{b_2}=4.8$ for $\phi=0.9$). This is in contrast with the results obtained for near-stoichiometric flames in a channel of width $d=20$, for which only symmetric solutions were obtained (see Fig.~\ref{fig:uf_m_d20}). The equivalence ratio must be increased up to $\phi=1.2$ to guarantee stable symmetric solutions at all flow rates when $d=80$. At this point a question arises: is there a critical equivalence ratio above which we could guarantee the symmetric solution for all channel widths and flow rates? As shall be shown in next section, for each couple of $Le_F$ and $Le_O$ there is indeed a critical equivalence ratio above which only symmetric flames are found. This critical equivalence ratio plays the role of the critical Lewis number in single-reactant formulations, in which, for cases with a Lewis number larger than unity, only symmetric flames can be found \cite{Ku11}.   


\section{Stability analysis}\label{sec:stability}
\noindent
In addition to the steady-state calculations described above, we carry out a linear stability analysis of the flame solutions perturbed in the form
\begin{align}
 \theta(x,y;t) &= \theta^0(x,y) + \epsilon \theta^1(x,y) e^{\lambda t}, \nonumber \\
  Y_i(x,y;t) &= Y^0_i(x,y) + \epsilon Y_i^1(x,y) e^{\lambda t}, \quad i=1,2, \nonumber 
  \end{align}
where $\lambda \in \mathbb{C}$, with $\lambda_R={\rm Re}(\lambda)$, and $\epsilon$ is a small amplitude. As demonstrated in \cite{Ku11}, infinitesimal perturbations of the flame propagation velocity can be excluded from the analysis without loss of generality. The superindex ``0'' is used to denote the steady flame base solution. Substitution of the above expressions into
(\ref{eq:T})-(\ref{eq:Y}) leads to the linearized problem: 
\begin{align}
\lambda \theta^1 &= -\sqrt{d}\{u_f+6\,my(1-y)\} \p{\theta^1}{x}+ \Delta \theta^1 + d \left(A \theta^1
+ B Y_2^1 + C Y_1^1 \right), \label{eq:pertur_T} \\
\lambda Y_i^1 &= -\sqrt{d}\{u_f+6\,my(1-y)\} \p{Y_i^1}{x}+ \dfrac{1}{Le_i}\Delta Y_i^1 - d \left(A \theta^1
+ B Y_2^1 + C Y_1^1 \right), \quad i=1,2,\label{eq:pertur_Y}
\end{align}
with 
\begin{align*}
A&=\p{\omega(\theta^0,Y_1^0,Y_2^0)}{\theta^0}=\dfrac{\beta^3 Y_1^0 Y_2^0}{2 {\cal L} s^2_L \left[1+\gamma(\theta_0-1) \right]^2}\exp\left\{ \dfrac{\beta(\theta^0-1)}{1+\gamma(\theta^0-1)} \right\}, \\[1ex]  
B&=\p{\omega(\theta^0,Y_1^0,Y_2^0)}{Y_2^0}=\dfrac{\beta^2 Y_1^0}{2 {\cal L} s^2_L}\exp\left\{ \dfrac{\beta(\theta^0-1)}{1+\gamma(\theta^0-1)} \right\}, \\[1ex]
C&=\p{\omega(\theta^0,Y_1^0,Y_2^0)}{Y_1^0}=\dfrac{\beta^2 Y_2^0}{2 {\cal L} s^2_L}\exp\left\{ \dfrac{\beta(\theta^0-1)}{1+\gamma(\theta^0-1)} \right\}. \
\end{align*}
Instead of directly solving the above linearized problem, we use the method developed in \cite{Ku11} and compute only the main eigenvalue, that is, the eigenvalue with the largest real part, in what follows denoted by $\lambda_R$. This is sufficient to determine the stability of the different solutions and is a simple to implement, low-cost method.

To study the emergence of non-symmetric shapes we perturb the symmetric base solution and solve the system of equations \eqref{eq:pertur_T}-\eqref{eq:pertur_Y} in half of the channel width with the boundary conditions
\begin{equation*}
\begin{aligned}
 x\to -\infty:&\quad \theta^1=Y_i^1=0,  &  x\to +\infty:&\quad \pl{\theta^1}{x}=\pl{Y_i^1}{x}=0, \\
 y=0 :&\quad \pl{\theta^1}{y}=\pl{Y_i^1}{y}=0, &  y=1/2 :&\quad \theta^1=Y_i^1=0,
\end{aligned}
\label{BCy_stability}
\end{equation*}
where $i=1,2$. The condition $\theta^1=Y_1^1=Y_2^1=0$ given at $y=1/2$ corresponds to the non-symmetric mode of the perturbation. When the largest eigenvalue is bigger than zero ($\lambda_R>0$) the symmetric solution is unstable and the flame acquires a non-symmetric shape. Another perturbation mode may correspond with the symmetric mode given by the boundary condition $\pl{\theta^1}{y}=\pl{Y_1^1}{y}=\pl{Y_2^1}{y}=0$ at $y=1/2$. In cases when this symmetric mode is stable, it always gives $\lambda_R=0$ in the computations. The reason is that the eigenfunction $(\theta^1,Y_1^1,Y_2^1)=(\pl{\theta^0}{x},\pl{Y_1^0}{x},\pl{Y_2^0}{x})$, which is symmetric with respect to the midplane, always exists with $\lambda=0$. For all the parameter combinations studied in the present work, we obtained $\lambda_R=0$ in the symmetric mode of the perturbation. The single-reactant model \cite{Ku11} predicts that values with $\lambda_R>0$ (including positive imaginary parts) only emerge for large enough Lewis numbers (typically $Le_1>4$). With this in mind, we have checked equivalence ratios as large as $\phi=7$ in the case $Le_O=2$, finding $\lambda_R=0$ as the largest eigenvalue in every computation, as expected. 

The stability analysis of the steady non-symmetric solution is also addressed in the present work. In some cases two different non-symmetric solutions can emerge close to a fold bifurcation and it is interesting to check the stability of both branches. To study the stability of non-symmetric solutions, we consider the entire channel width and the (only possible) perturbation mode $\pl{\theta^1}{y}=\pl{Y_1^1}{y}=\pl{Y_2^1}{y}=0$ at $y=1$.

\begin{figure}[t]
\begin{center}
\includegraphics[width=0.49\textwidth]{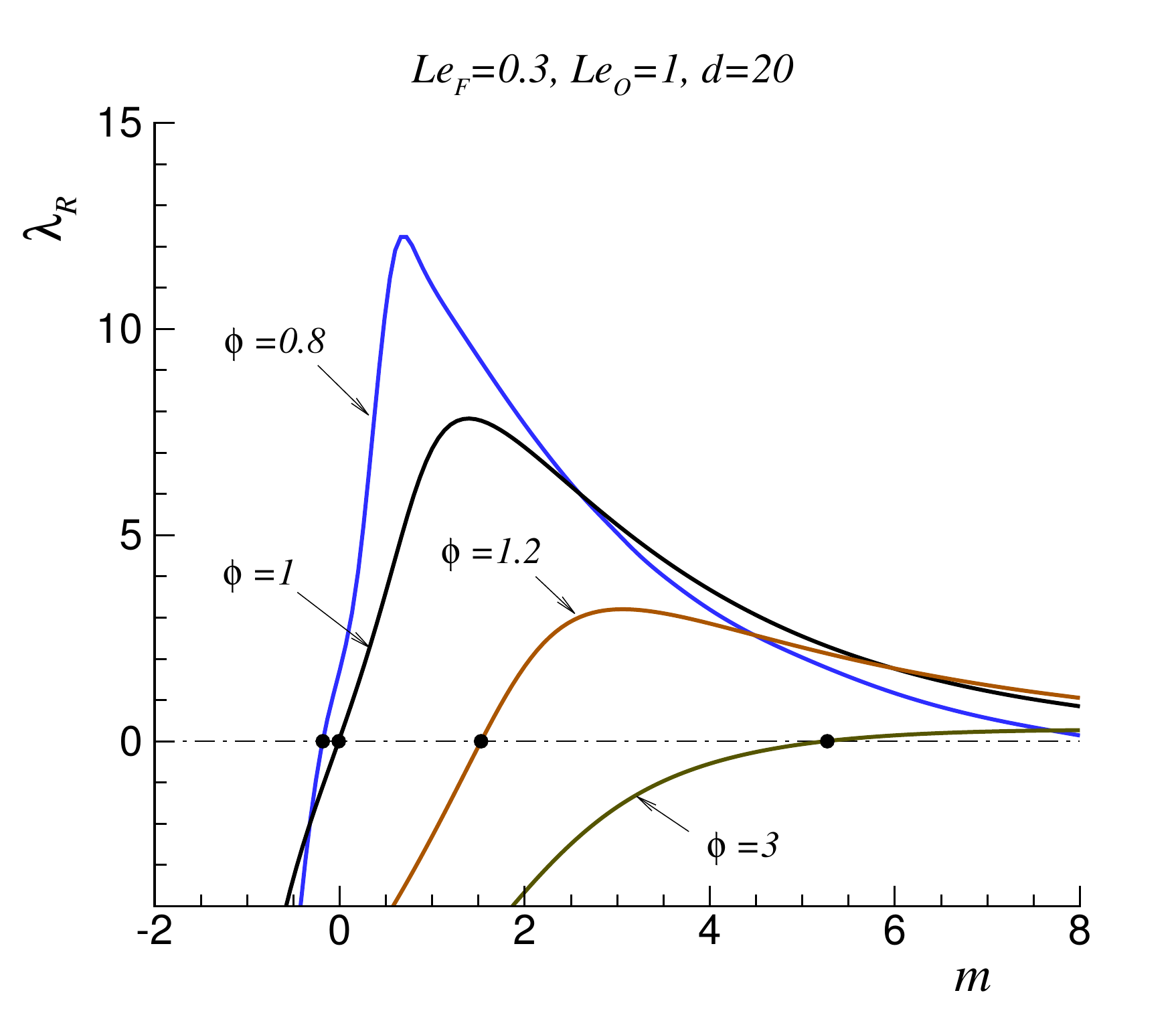}
\includegraphics[width=0.49\textwidth]{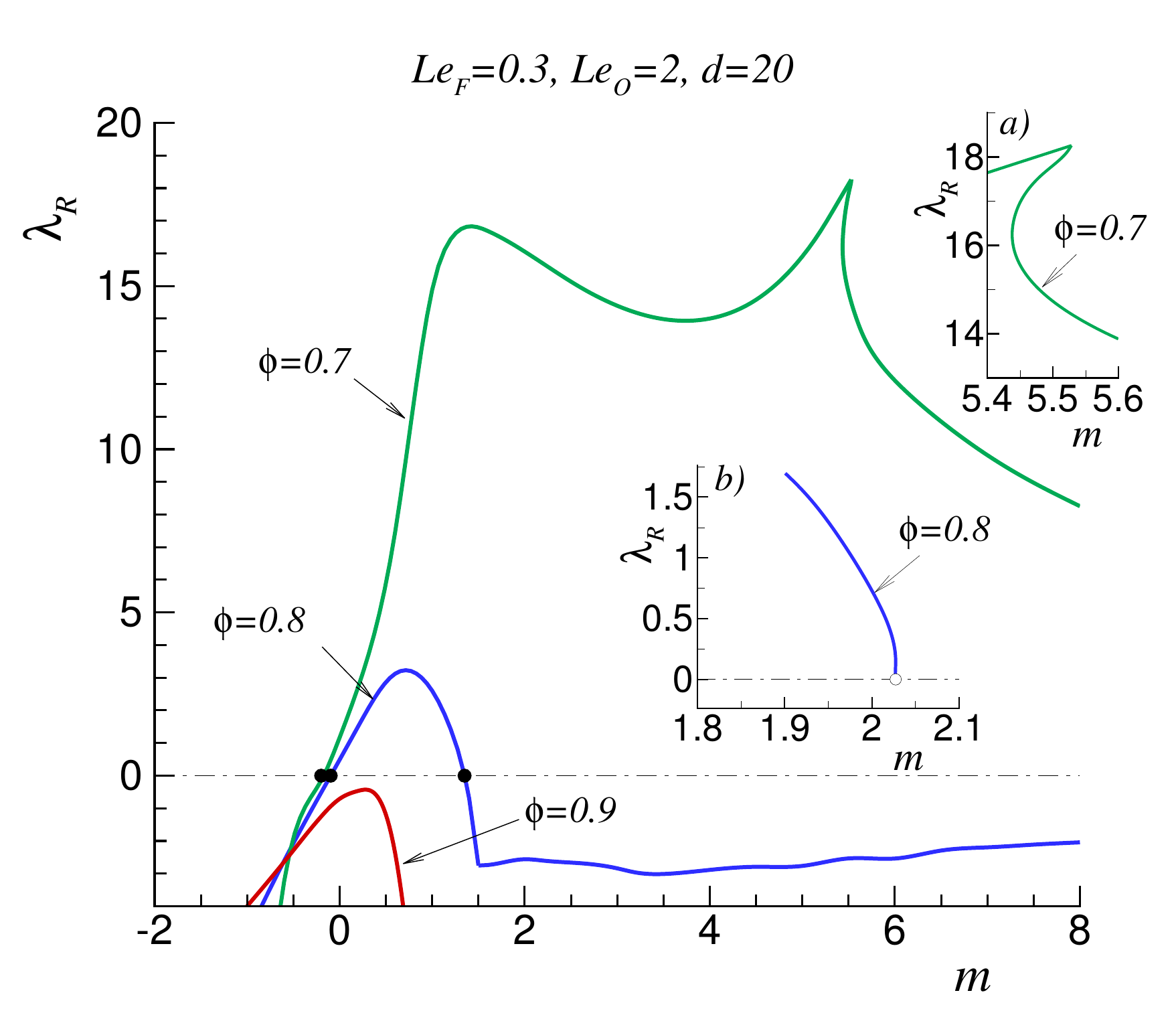}
\caption{The growth rate $\lambda_R$ of the steady symmetric solution plotted as a function of the flow rate for several equivalence ratios; the inset $a)$ blows up the region of multi-valued symmetric solutions for $\phi=0.7$; the inset $b)$ gives $\lambda_R$ with $m$ for the steady non-symmetric solution $\phi=0.8$ close the fold bifurcation point. Calculated for $Le_F=0.3$, $Le_O=1$ and $d=20$ (left) and with $Le_O=2$ (right).}
\label{fig:lbr_m_d20}
\end{center}
\end{figure}

In Fig.~\ref{fig:lbr_m_d20} we plot the calculated real part of the main eigenvalue $\lambda_R$ versus the flow rate for the cases with $Le_O=1$ (left) and $Le_O=2$ (right). The figure corresponds to the stability analysis of the symmetric solution for $d=20$ and validates the results of the previous section. For example, the symmetric solution becomes unstable ($\lambda_R>0$) for mass flow rates $m>m_{b_1}$, in agreement with the steady-state calculations. For $Le_O=1$ we always find a range of flow rates where the symmetric solution is unstable. The mass flow rate at the bifurcation point $m_{b_1}$ grows with the equivalence ratio. For $Le_O=2$ a critical equivalence ratio exists ($\phi \approx 0.89$). The inset $a)$ blows up the region of multi-valued unstable symmetric solutions for $\phi=0.7$.

\begin{figure}[t]
\begin{center}
\includegraphics[width=0.49\textwidth]{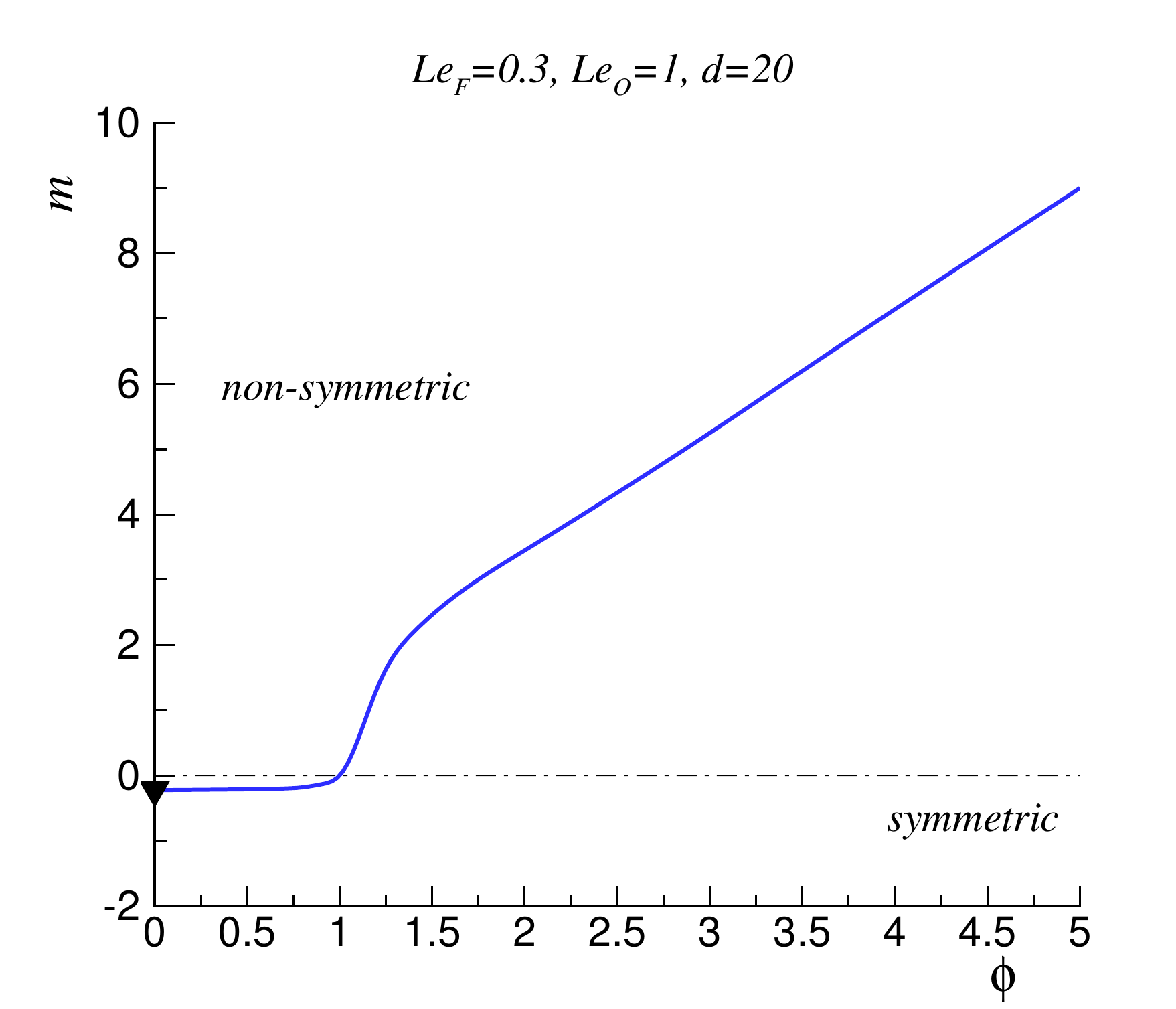}
\includegraphics[width=0.49\textwidth]{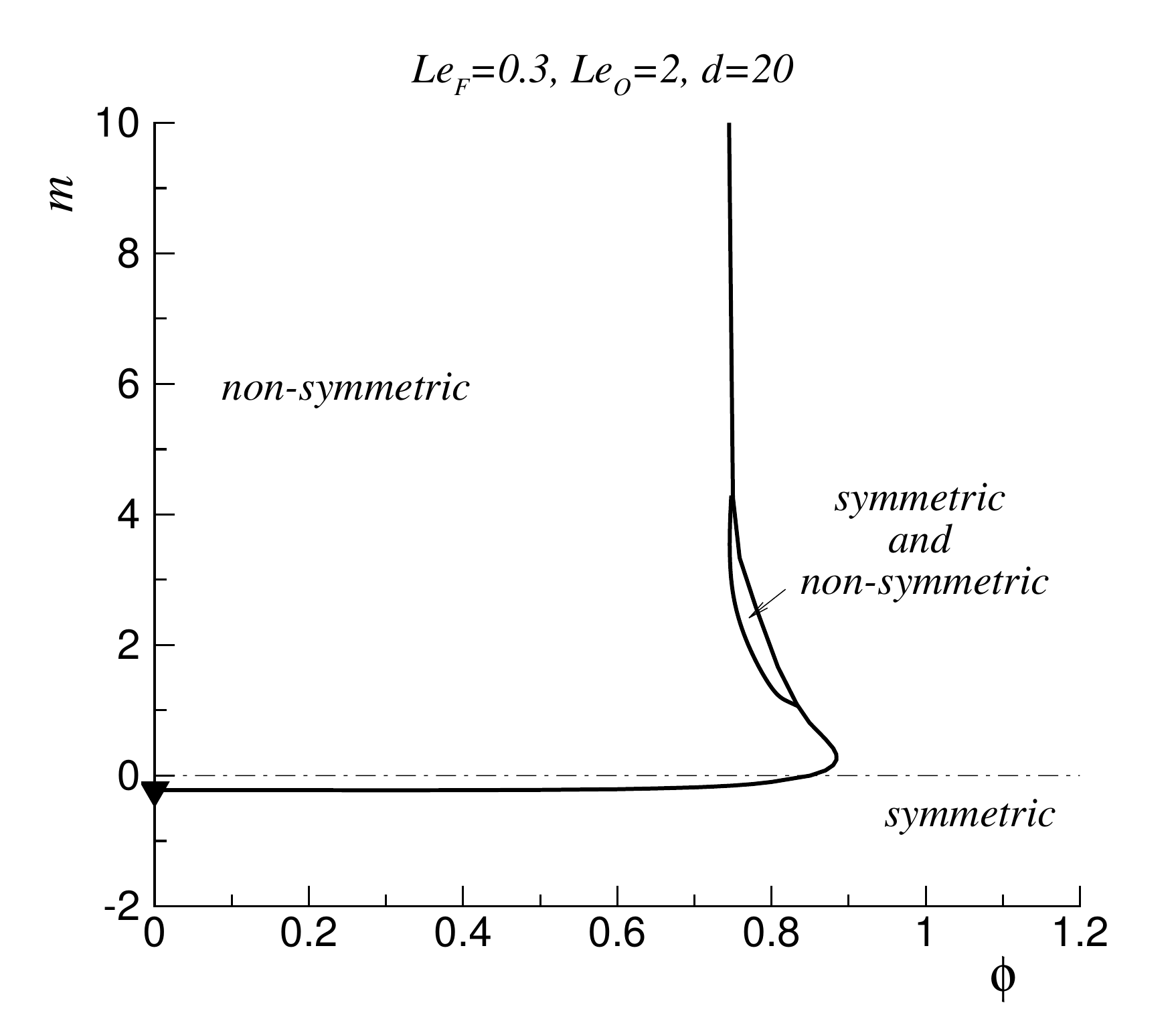}
\caption{Stability map delineating the regions with symmetric and non-symmetric solutions in the $m$ vs. $\phi$ parametric space; the symbol $\blacktriangledown$ indicates the calculation for a single specie given in \cite{Ku11}. Calculated for $Le_F=0.3$, $Le_O=1$ and $d=20$ (left) and with $Le_O=2$ (right).}
\label{fig:stab_m_d20}
\end{center}
\end{figure}

In Fig.~\ref{fig:stab_m_d20} we show a complete stability map in the $m$ vs. $\phi$ parametric space delineating the stable regions of the symmetric and the non-symmetric solutions. In the case $Le_O=1$ (left) we find a linear dependence between the values $m$ and $\phi$ that separate symmetric and non-symmetric flames in the fuel-rich zone, indicating that non-symmetric solutions will appear for any equivalence ratio if $m$ is sufficiently large. On the other hand, the case $Le_O=2$ (right) shows a nose-like curvature with an asymptotic value of $\phi$ above which the flame will always be symmetric independently of the value of $m$. The asymptotic behavior of the stability map is linked with the differential diffusion and is somehow related to the effective Lewis number of the mixture. The multiplicity of solutions is, nevertheless, associated with the preferential diffusion in near-stoichiometric flames. In this case, the concentration of both reactants is small in the reactive-diffusive part of the flame and the less diffusive specie (the oxidizer in this case) finds more difficulties to reach the flame. The convective transport still supplies the necessary amount of oxidizer near the axis of the channel but not near the walls, where the velocity is small. The oxidizer reaches the near-wall region by transverse diffusion, creating a region with $Y_O$ below the equilibrium value $(1-\phi)/\phi$ that forces the flame to adopt the symmetric shape shown in Fig.~\ref{fig:YO_d20} for large mass flow rates.

It is worth mentioning that the limit of very lean flames studied by Kurdyumov \cite{Ku11} (and indicated with the symbol $\blacktriangledown$ in the figure) is recovered for flames with an equivalence ratio not necessarily very lean, at around $\phi \simeq 0.6$. Of course, the limit $\phi \to 0$ must be understood as the limit of very lean flames, where the oxidant is in excess and the reaction rate is governed by the fuel concentration.

We consider now the stability of the non-symmetric branch for the case $Le_O=2$ and with $\phi=0.8$ presented before in Fig.~\ref{fig:uf_m_d20}. Above the fold bifurcation point, marked with symbol $\circ$, $\lambda_R=0$ and the branch is stable. Below this point, $\lambda_R>0$ and the branch (marked with dotted line) becomes intrinsically unstable. The growth rate of this branch close to the fold bifurcation point is plotted in the inset \textcolor{red}{$b)$} of Fig.~\ref{fig:lbr_m_d20} (right), indicating that the branch between points $m_{b_2}$ and $m_{b_3}$ is unstable.

\begin{figure}[!ht]
\begin{center}
\includegraphics[width=0.49\textwidth]{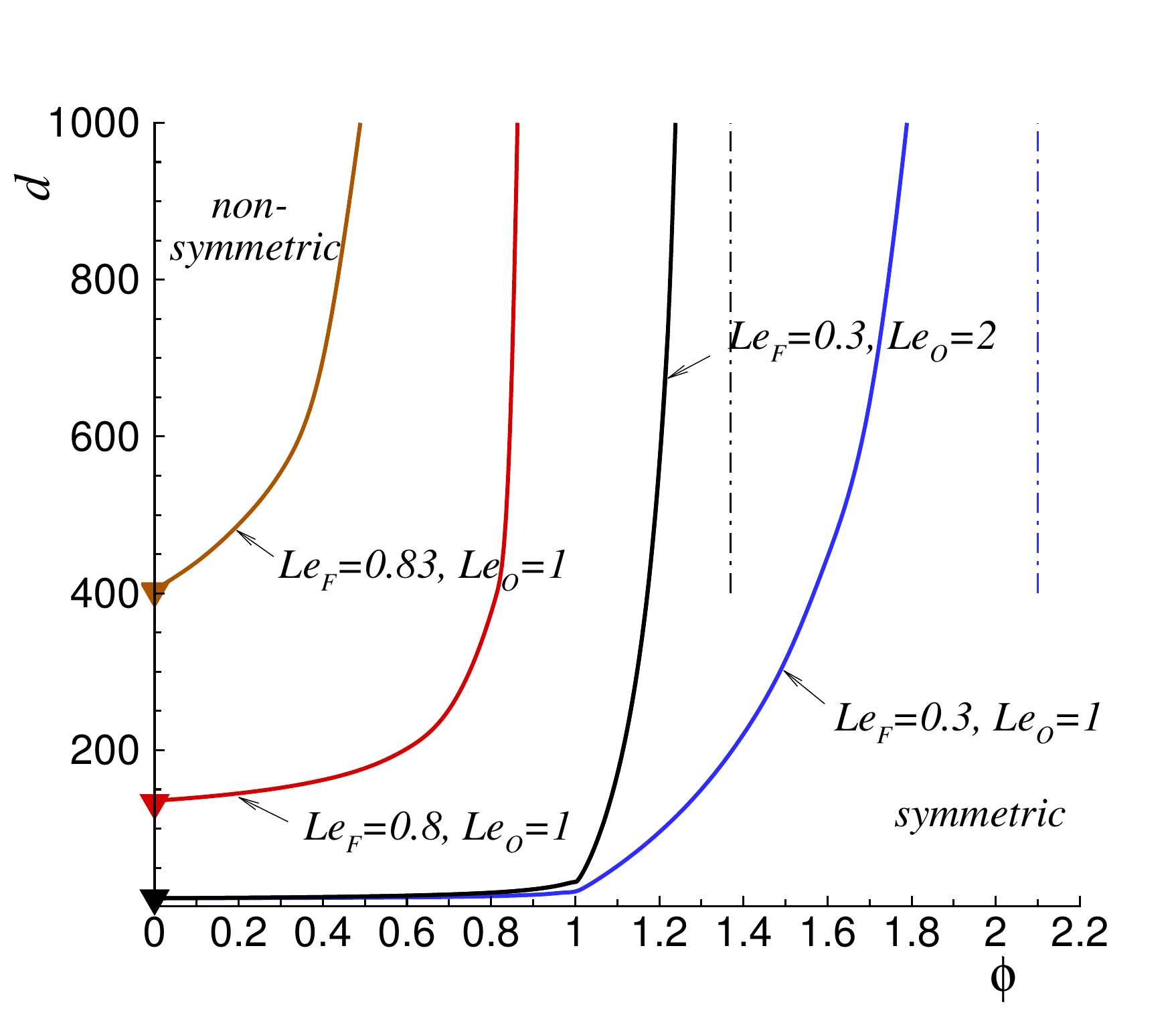}
\caption{Stability map delineating the regions with symmetric and non-symmetric solutions in the $d$ vs. $\phi$ parametric space. The triangle symbols indicate calculations for a single specie given in \cite{Ku11}, where $d_c=11.2, 132.8$ and 402.9 for $Le_F=0.3, 0.8$ and 0.83, respectively. The asymptotes (dot-dashed lines) were calculated from the linear stability analysis of the planar flame, with $\beta=10$ and $\gamma=0.8$.}
\label{fig:dc_m}
\end{center}
\end{figure}

Qualitatively, the size of the channel does not change the stability behavior of the solution. As commented before, it only modifies the range of equivalence ratios and flow rates where the non-symmetric solutions emerge. In Fig.~\ref{fig:dc_m} we plot the stability map for $m=0$ in the $d$ vs. $\phi$ parametric space, delineating again the stable regions of the symmetric and the non-symmetric solutions. These curves were calculated in a different manner than those shown in Fig.~\ref{fig:lbr_m_d20}. To determine the critical value of $d$ we imposed $\lambda=0$ in Eqs.~\eqref{eq:pertur_T}-\eqref{eq:pertur_Y} and non-symmetric perturbations of the type $\theta(x)=\theta^0(x)+\epsilon \theta^1(x) \cos(\pi y)$ and $Y_i(x)=Y_i^0(x)+\epsilon Y_i^1(x)\cos(\pi y)$, with $i=1,2$. The resulting one-dimensional eigenvalue problem becomes more computationally affordable \cite{Ku11}. Fig.~\ref{fig:dc_m} shows that for each studied combination of $Le_F$ and $Le_O$ the flame is non symmetric above and to the left of the corresponding curve. For large values of $d$ the limiting equivalence ratio tends asymptotically to a critical value, that we denote by $\phi_c$. This is the critical equivalence ratio above which we can guarantee the symmetric solution for all channel widths (for $m=0$). In this limit, when $d$ is sufficiently large, the problem of a flame propagating in an adiabatic channel is equivalent to that of a planar flame and the stability of both problems should give identical results. For completeness, we present in Appendix \ref{appendix} the global stability analysis of a planar flame using a two-reactant model. Results for the value of $\phi_c$ calculated for the planar problem are plotted in Fig.~\ref{fig:dc_m} using dot-dashed vertical lines for cases with $Le_F=0.3$ and $Le_O=1$, and $Le_O=2$. These values compare well with the asymptotic tendency of the curves. For $m>0$ the computations become more time demanding and results are not presented here. However, it is clear that $\phi_c$ must increase up to a value dictated by the condition given by the relative values of $Le_F$ and $Le_O$, or alternatively, by the value of the effective Lewis number of the mixture. 

The results of the present stability analysis are in agreement with those of the single-reactant model given in \cite{Ku11} (and marked with triangle symbols in the figure for different values of $Le_F$). For sufficiently large values of $d$, the stability problem of the single-reactant model should again give the same results than those of the planar flame. Lasseigne et al. \cite{LaJaJa99} addressed the stability analysis of a planar flame with finite activation energy and showed that flames destabilize when the Lewis number is smaller than a critical value close to unity, that for $\beta=10$ and $\gamma=0.8$ corresponds to $Le_F\approx 0.84$. Fig.~\ref{fig:dc_m} clearly illustrates this tendency.


\section{Conclusions}
Steady-state calculations and linear stability analysis are employed to investigate the effect that stoichiometry variations have on the break of symmetry for flames propagating in narrow adiabatic channels. We use a two-reactant constant-density formulation in which the fuel an oxidizer mass fraction equations are solved together with the temperature equations. The fuel and oxidizer Lewis numbers are chosen as $Le_F=0.3$ and $Le_O=2$, characteristics of very lean and very rich hydrogen-oxygen systems, respectively, so that varying only the parameter $\phi$ would cover a wide range of possible behaviors (or possible effective Lewis numbers).

The computations show that lean mixtures suffer from differential-diffusion induced instabilities, as dictated by the value of $Le_F<1$, and that increasing the flow rate contributes to the flame destabilization toward non-symmetric shapes, in agreement with results reported before in \cite{Ku11} for a single-reactant model. With use of the two-reactant model the stability of the symmetric flame shape is given by a weighted average of $Le_F$ and $Le_O$ and therefore depends on the equivalence ratio. The new phenomena observed are:
  \begin{itemize}
  \item the existence of a critical equivalence ratio above which the symmetric flame is stable for any value of the flow rate for oxidizers with large diffusivity, as shown in Fig~\ref{fig:dc_m},
  \item the stabilization of the symmetric flame with the increase of both the flow rate and the equivalence ratio,
  \item the existence of multiple stable solutions.
  \end{itemize}
The last two phenomena emerge, in particular, for near-stoichiometric mixtures. At this condition, when the concentration of the two reactants is comparably small in the reaction zone, the preferential diffusion effect becomes very important and the species with the slower diffusion has difficulties to reach the flame. In those cases, an increase of the flow rate promotes the stabilization of the symmetric shape and the appearance of multiplicity of solutions.

The paper demonstrates that previous investigations assuming the limit of lean flames for a single species give satisfactory results even for equivalence ratios not necessarily very lean, at around $\phi \simeq 0.6$.



\section*{Funding}
The research leading to these results has received funding from the European Union's Horizon 2020 Programme (2014-2020) and from Brazilian Ministry of Science, Technology and Innovation through Rede Nacional de Pesquisa (RNP) under the HPC4E Project (www.hpc4e.eu), grant agreement number 689772, and from the projects ENE2015-65852-C2-1-R and ENE2015-65852-C2-2-R (MINECO/FEDER).
\appendix
\section{The planar flame stability analysis}\label{appendix}

To the authors' knowledge, no analysis of the linear stability of two-reactant flames with finite activation energy can be found the literature. Joulin and Mitani's results \cite{JoMi81} are based on the high activation energy asymptotic (HAEA) limit and the flame is modeled with the classical assumption of an infinitely thin discontinuity. To cover that gap, we briefly present here the formulation of the two-reactant stability problem and calculate the critical equivalence ratio, $\phi_c$, defined as the value of $\phi$ above which the planar flame is intrinsically stable. The obtained values are used to draw the asymptotes in Fig.~\ref{fig:dc_m} and should correspond to the critical equivalence ratio for channels sufficiently wide. In the computations we found remarkable discrepancies with the asymptotic values of Joulin and Mitani. For example, the critical equivalence ratio predicted within the HAEA limit for $Le_F=0.3$ and $Le_O=1$ was about $\phi_c \approx 1.3$ and for $Le_F=0.3$ and $Le_O=2$ was about $\phi_c \approx 0.91$. Our calculations with $\beta=10$ give values much larger than those calculated with infinite activation energy, as indicated below. It is well-known that the activation energy must increase up to really large values ($\beta > 60$) for the asymptotic results to be valid, as clarified in the stability of single-reactant flames with finite activation energy addressed in \cite{LaJaJa99}.  

\begin{figure}[!ht]
\begin{center}
\includegraphics[width=0.49\textwidth]{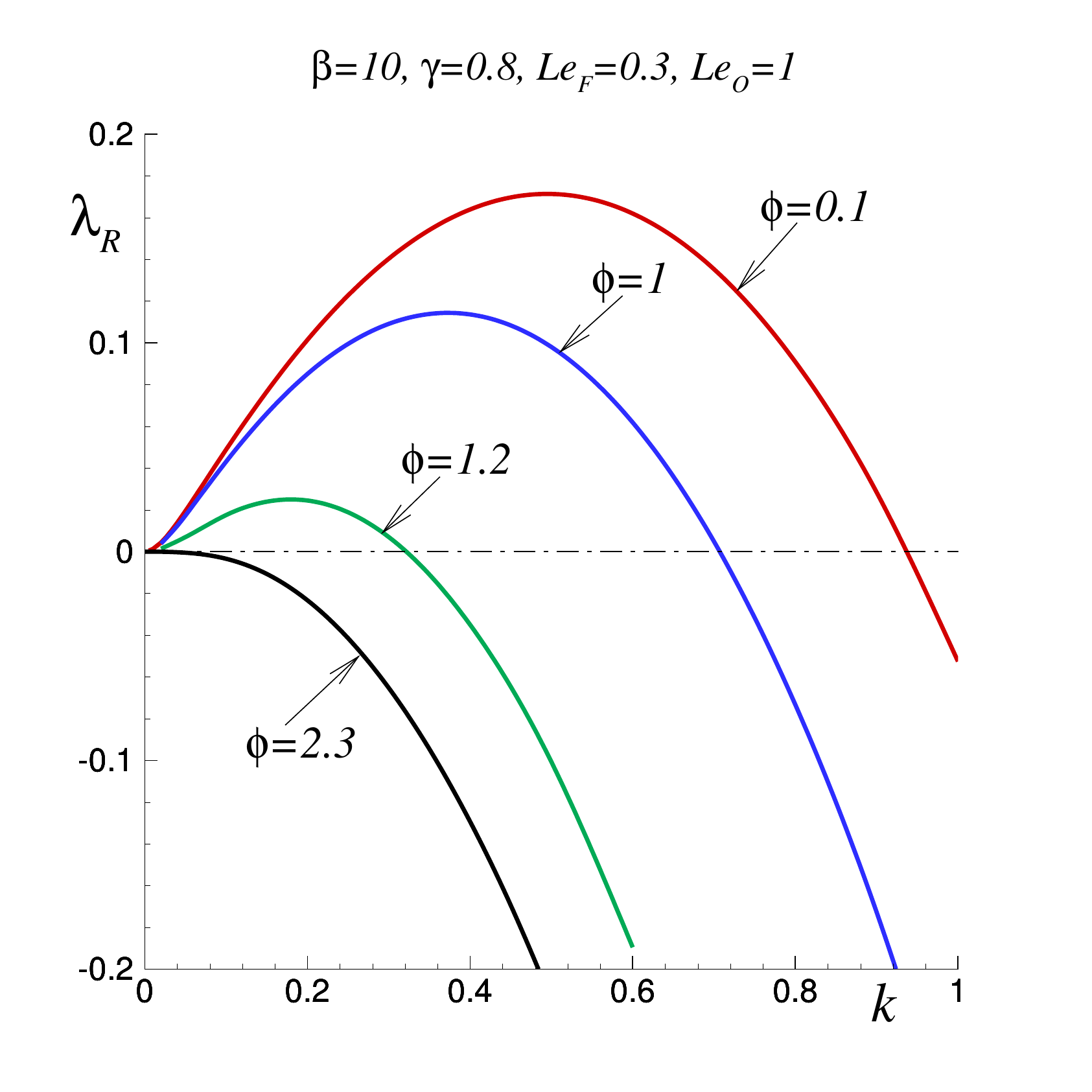}
\includegraphics[width=0.49\textwidth]{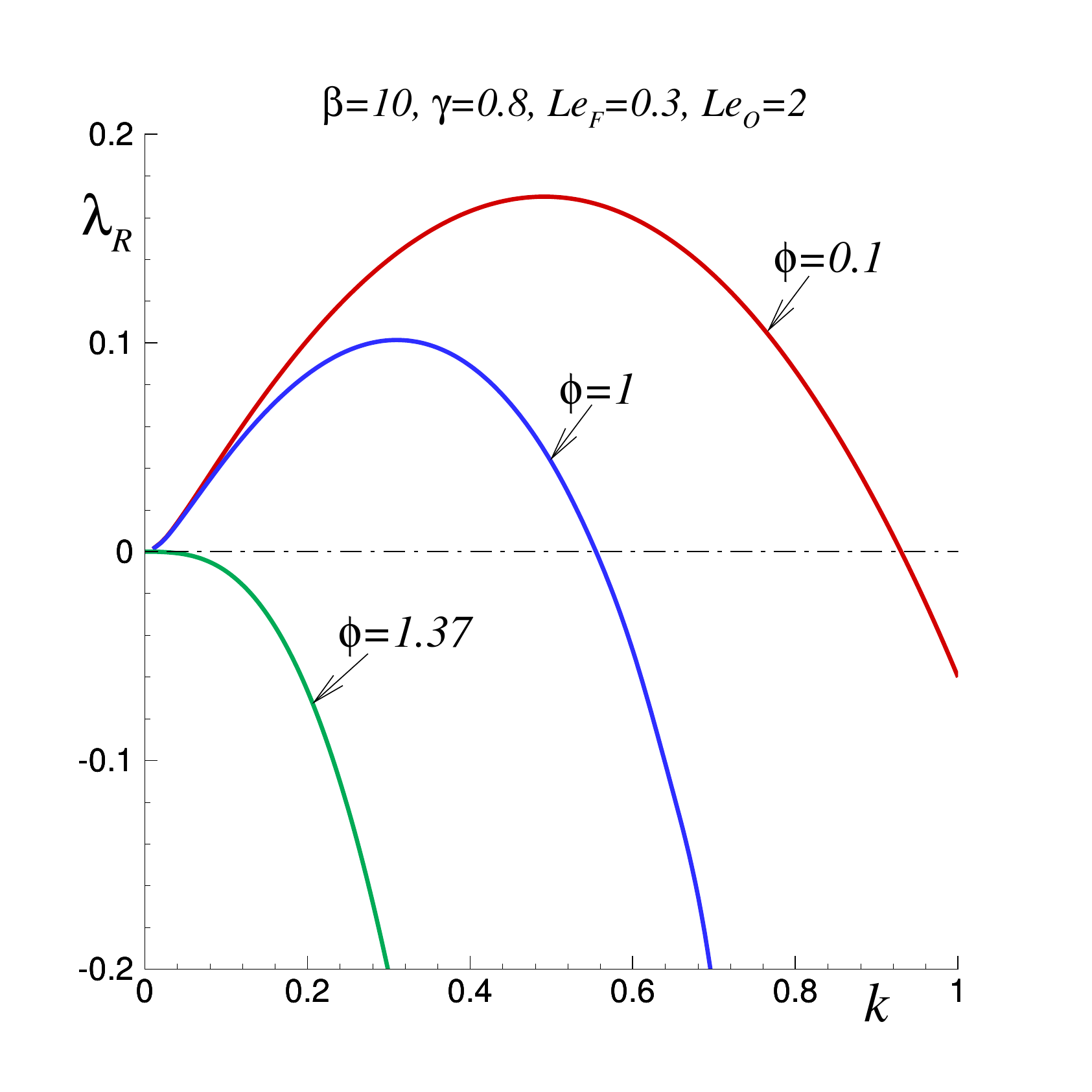}
\caption{Planar flame stability curves: the variation with the wave number $k$ of the growth rate $\lambda_R$. Calculated for $\beta=10$, $\gamma=0.8$ and $Le_F=0.3$ with $Le_O=1$ (left) and with $Le_O=2$ (right).}
\label{fig:lb_k}
\end{center}
\end{figure}

In this appendix we calculate the critical equivalence ratio as the value at which the main eigenvalue $\lambda_R$ of the stability problem becomes negative for all range of wave numbers $k$. When the planar problem \eqref{eq:planar} is perturbed as usual in the form  
\begin{align}
 \theta(\xi;\tau) &= \theta^0(\xi) + \epsilon \theta^1(\xi) e^{\lambda \tau + j k \eta}, \nonumber \\
  Y_i(\xi;\tau) &= Y^0_i(\xi) + \epsilon Y_i^1(\xi) e^{\lambda \tau + j k \eta}, \quad i=1,2, \nonumber 
  \end{align}
where $j=\sqrt{-1}$, $\eta=y\sqrt{d}$ and $\tau=t\sqrt{d}$, it leads to the linearized problem
\begin{align*}
\lambda \theta^1 &= -\der{\theta^1}{\xi}+ \dd{\theta^1}{\xi} - k^2\theta^1 + \left(A \theta^1 + B Y_2^1 + C Y_1^1 \right), \label{eq:ppertur_T} \\
\lambda Y_i^1 &= -\der{Y_i^1}{\xi}+ \dfrac{1}{Le_i}\left(\dd{Y_i^1}{\xi} - k^2 Y_i^1 \right)- \left(A \theta^1 + B Y_2^1 + C Y_1^1 \right), \quad i=1,2,
\end{align*}
with the expression for $A$, $B$ and $C$ given in Section \ref{sec:stability}. The problem has to be complemented with the boundary conditions
\begin{equation*}
 \xi\to -\infty:\, \theta^1=Y_1^1=Y_2^1=0,  \qquad  \xi\to +\infty:\, \dl{\theta^1}{\xi}=\dl{Y_1^1}{\xi}=\dl{Y_2^1}{\xi}.
\label{BCx_pstability}
\end{equation*}
An example of the calculated values of $\lambda$ is given in Fig.~\ref{fig:lb_k}, where the main eigenvalue $\lambda_R$ is plotted versus the wave number $k$. For the case $Le_F=0.3$ and with $Le_O=1$ it corresponds to $\phi_c=2.1$ and for the case with $Le_O=2$ the value is $\phi_c=1.37$. These values correspond to the asymptotes (dot-dashed lines) of Fig.~\ref{fig:dc_m}.  

\end{document}